\documentclass[sigconf]{acmart}
\AtBeginDocument{%
  }
\usepackage{microtype}
\usepackage{graphicx}
\usepackage{array}
\usepackage{booktabs}
\usepackage{amsmath}
\usepackage{mathtools}
\usepackage{multirow}
\usepackage[capitalize,noabbrev]{cleveref}
\usepackage[table]{xcolor}
\usepackage{makecell}
\usepackage{adjustbox}
\usepackage{pifont}
\usepackage{afterpage}
\usepackage{colortbl}
\usepackage{graphicx}
\definecolor{hdrgray}{gray}{0.90}
\definecolor{allcol}{HTML}{EDF2FA}
\definecolor{oursrow}{HTML}{E2EDF8}
\definecolor{dimtext}{gray}{0.42}
\definecolor{heat}{HTML}{E4572E}
\newcommand{\nametrap}{\texttt{NameTrap}}
\ExplSyntaxOn
\NewDocumentCommand{\mdl}{m}{
  \tl_set:Nn \l_tmpa_tl {#1}
  \tl_replace_all:Nnn \l_tmpa_tl {-} {-\allowbreak}
  \texttt{\l_tmpa_tl}
}
\ExplSyntaxOff
\hypersetup{breaklinks=true}

\setcopyright{none}
\acmConference[WWW '27]{The ACM Web Conference 2027}{May 10--14, 2027}{Dublin, Ireland}
\acmYear{2027}
\copyrightyear{2027}

\author{Chieh-Yuan Chang$^\dagger$, Guan-Wei Chen$^\dagger$, Chia-Yi Hsu$^\dagger$, Chih Yang$^\dagger$, Chia-Mu Yu$^\dagger$, Chun-Ying Huang$^\dagger$}
\affiliation{\institution{$^\dagger$National Yang Ming Chiao Tung University}\country{Taiwan}}
\begin{document}

\title{\texorpdfstring{\nametrap}{NameTrap}: Steering AI Coding Agents Toward Security Vulnerabilities via Variable Names}

\begin{abstract}
AI coding agents routinely clone repositories, install packages, and read source code obtained from the Web, so publicly hosted code becomes part of the context that shapes the code they write. We show that this Web-borne context can be weaponized without any explicit instruction. We present \nametrap, a source-code poisoning attack that only renames identifiers in third-party Python files to steer AI coding agents toward attacker-targeted Common Weakness Enumeration (CWE) vulnerabilities. An attacker-side coding agent iteratively aligns identifiers with a CWE-specific naming convention and confirms through local probe completions that the renamed file elicits the target weakness, while the released bait contains neither instructions nor vulnerable statements. We evaluate \nametrap{} across six AI coding agents, ten underlying language models, and four vulnerability categories: CWE-78, CWE-79, CWE-327, and CWE-502. Starting from files whose clean versions never yield the targeted CWE, identifier-level manipulation achieves overall attack success rates of 22.5\% to 56.0\% across the six agents, reaching 75.0\% for certain agent--model configurations, and substantially outperforms random renaming, neutral renaming, and the prior identifier-poisoning attack XOXO. The attack remains effective under Traditional Chinese prompts and transfers across models, while a security-reminder defense fails to consistently prevent vulnerability introduction. Moreover, three AI-based security analyzers and four static analyzers report none of the targeted weaknesses in the poisoned files. These findings expose an overlooked attack surface of agents that consume Web content: seemingly harmless variable names in untrusted, Web-hosted code can compromise the security of downstream software.
\end{abstract}

\ccsdesc[500]{Security and privacy~Software security engineering}
\ccsdesc[300]{Computing methodologies~Artificial intelligence}

\keywords{AI coding agents, source-code poisoning, identifier renaming, software supply chain, LLM agent security}

\maketitle

\section{Introduction}
\label{sec:introduction}

Large language model (LLM)-based coding agents are changing how software is developed. Beyond generating individual snippets, they inspect repositories, execute tools, edit multiple files, and iteratively test their changes~\cite{yao2023react,jimenez2024swebench,yang2024sweagent,wang2025openhands,xia2024agentless}. Much of the context they act on comes from the Web: agents clone GitHub repositories, install dependencies from public package registries, and retrieve documentation and example code through Web tools. Repository-level completion research shows that such cross-file context directly shapes subsequent code generation~\cite{liu2024repobench,zhang2023repocoder}. Code hosted on the open Web has therefore become part of the decision-making context of coding agents, and its integrity is a Web security problem: a single public file can influence every agent that reads it. This paper studies how an adversary can exploit this channel, addressing the security of AI agents that consume Web content.

This exposure differs from conventional software supply-chain attacks, in which a malicious dependency itself implements unwanted behavior~\cite{zimmermann2019npm}. In AI-assisted development, an adversary can instead influence the \emph{new} code that an agent writes by altering material the agent reads during a legitimate task. The risk is amplified because secure code generation is difficult even without an adversary: AI assistants frequently produce insecure code~\cite{pearce2022asleep,perry2023users}, as confirmed by security benchmarks~\cite{peng2025cweval,vero2025baxbench,bhatt2024cyberseceval2}, including repository-level ones~\cite{shen2025secrepobench,wang2026realsec}. We ask whether an attacker can deliberately raise the likelihood of a \emph{particular} weakness through the context the agent reads.

LLM agents are known to be redirectable through lower-trust content, via indirect prompt injection~\cite{greshake2023indirect,zhan2024injecagent,debenedetti2024agentdojo}, memory and knowledge poisoning~\cite{chen2024agentpoison,zou2025poisonedrag}, and broader attacks on tools and actions~\cite{zhang2025asb,ruan2024toolemu}. For coding agents, repository artifacts such as documentation and agent rule files can carry injected instructions~\cite{li2026repoguardbench,zhu2026beyond}. These attacks rely on explicit natural-language directives. Source-code identifiers offer a subtler channel: names convey roles, expected values, and trust assumptions, yet renaming them leaves execution semantics unchanged. Code models exploit identifier semantics explicitly~\cite{guo2021graphcodebert,wang2021codet5}, semantics-preserving renaming changes model outputs~\cite{yefet2020adversarial,yang2022alert,jha2023codeattack,wen2025variable}, and XOXO showed that such changes can poison the cross-origin context of AI coding assistants~\cite{storek2026xoxo}. This raises a concrete question for autonomous coding workflows: can an innocuous-looking identifier change in a file obtained from the Web steer an agent into implementing a \emph{specific} vulnerability?

We answer this question with \nametrap, an agent-generated source-code poisoning attack that renames identifiers in third-party Python files to induce attacker-targeted Common Weakness Enumeration (CWE) vulnerabilities during downstream code generation. The attacker publishes a file on a public code-hosting platform but cannot change the victim agent's system instructions, the developer's task, the model, or the agent framework. An attacker-side coding agent iteratively applies CWE-conditioned naming conventions to an incomplete file and probes each variant with a local completion and security scan. The released bait retains only admissible identifier changes; it contains neither instructions nor the vulnerable completion produced by the probe. The weakness materializes only when a separate victim agent, performing a legitimate task, interprets the poisoned names.

We evaluate \nametrap{} on Python files from 31 popular GitHub projects with six coding agents (OpenCode, Claude Code, Codex, mini-swe-agent, Aider, and OpenHands), ten LLMs, and four CWEs: OS command injection (CWE-78), cross-site scripting (CWE-79), broken or risky cryptography (CWE-327), and unsafe deserialization (CWE-502). All experiments start from screened clean files that never yield the targeted CWE in ten baseline completions. \nametrap{} reaches overall attack success rates (ASRs) of 22.5\%--56.0\% across agents and up to 75.0\% for individual agent--model configurations, whereas random renaming, neutral renaming, and XOXO are far less effective. The attack persists under Traditional Chinese prompts, transfers across models, and is not consistently suppressed by an explicit security reminder; meanwhile, seven AI-based and static analyzers detect none of the targeted weaknesses in the bait files. In summary, we make the following contributions:
\begin{itemize}
    \item \textbf{Identifier-only, agent-generated source poisoning.} We define a threat model and an iterative, probe-verified bait-construction process that uses CWE-conditioned identifier renaming, rather than explicit instructions or injected vulnerable statements, to steer downstream code generation.
    \item \textbf{Systematic evaluation of targeted vulnerability induction.} We evaluate six coding agents, ten LLMs, and four CWE categories against screened clean baselines, and compare against random renaming, neutral renaming, and XOXO.
    \item \textbf{Transferability, defense, and detection analysis.} We measure cross-model transfer and prompt-language effects, and show that a prompt-level security reminder and seven pre-completion vulnerability analyzers do not mitigate the attack.
\end{itemize}

\section{Related Work}
\label{sec:related-work}

\textbf{Security of AI-generated code.} Functional-looking code produced by AI assistants is often insecure: Pearce et al.\ found CWE-relevant weaknesses in GitHub Copilot suggestions, and Perry et al.\ showed that developers using AI assistants wrote less secure code while being more confident in it~\cite{pearce2022asleep,perry2023users}. Benchmarks now assess security jointly with functionality, from CWE-oriented and end-to-end backend tests~\cite{peng2025cweval,vero2025baxbench} and broad cybersecurity suites~\cite{bhatt2024cyberseceval2} to repository-level settings~\cite{shen2025secrepobench,wang2026realsec}. The latter build on repository-aware completion and issue-resolution benchmarks~\cite{liu2024repobench,zhang2023repocoder,jimenez2024swebench} and on agent scaffolds that read, edit, and test entire repositories~\cite{yao2023react,yang2024sweagent,wang2025openhands,xia2024agentless}. He and Vechev further showed that code LLMs can be steered toward or away from vulnerable code through learned prefixes~\cite{he2023sven}. These works measure how often insecure code arises; \nametrap{} measures how far an adversary can raise the likelihood of a \emph{specific} CWE through the context an agent reads, relative to screened clean baselines.

\textbf{Prompt injection and context poisoning of LLM agents.} Indirect prompt injection embeds instructions in Web pages, documents, and other lower-trust content consumed by LLM-integrated applications~\cite{greshake2023indirect,liu2024formalizing}. InjecAgent, AgentDojo, ASB, and ToolEmu benchmark its effect on tool-using agents~\cite{zhan2024injecagent,debenedetti2024agentdojo,zhang2025asb,ruan2024toolemu}, EIA demonstrates it against generalist Web agents~\cite{liao2025eia}, and AgentPoison and PoisonedRAG poison agent memory and retrieval corpora~\cite{chen2024agentpoison,zou2025poisonedrag}. For coding agents, instructions planted in repository files, rule files, and documentation can trigger harmful shell commands~\cite{liu2025yourshell,li2026repoguardbench}, with susceptibility depending on how users invoke the agent~\cite{zhu2026beyond}. Defenses such as the instruction hierarchy~\cite{wallace2024hierarchy} and structured queries~\cite{chen2025struq} separate trusted instructions from untrusted data. All of these settings assume that the adversarial content is, or contains, an instruction. \nametrap{} carries no instruction: its payload is a set of identifiers that remain valid program text and merely suggest an implementation convention, a channel that instruction--data separation does not directly target.

\textbf{Identifier semantics and adversarial code transformations.} Code models exploit developer-assigned names: \mdl{CodeT5} uses identifier-aware pretraining, and \mdl{GraphCodeBERT} models data flow among variables~\cite{wang2021codet5,guo2021graphcodebert}. Consequently, semantics-preserving transformations, notably variable renaming, can flip model predictions~\cite{yefet2020adversarial,yang2022alert,jha2023codeattack,wen2025variable} and degrade the robustness of code generation~\cite{wang2023recode}. These studies target prediction errors on fixed tasks; \nametrap{} instead asks whether renamed identifiers cause an agent to \emph{write new code} containing an attacker-chosen CWE.

\textbf{Poisoning code models and the software supply chain.} Training-time poisoning can make completion models suggest insecure code~\cite{schuster2021autocomplete,aghakhani2024trojanpuzzle,yan2024codebreaker}, with triggers hidden in adversarial identifiers~\cite{yang2024afraidoor} or code style~\cite{tran2026poisonstyle}. Open-source ecosystems distributed over the Web amplify the reach of a single malicious package~\cite{zimmermann2019npm}, and code LLMs add new supply-chain risks such as hallucinated package names~\cite{spracklen2025package}. \nametrap{} needs no access to training data or model parameters and ships no malicious code; it acts at inference time through ordinary source files. The closest work, XOXO~\cite{storek2026xoxo}, introduced cross-origin context poisoning of AI coding assistants via semantics-preserving changes, including identifier renaming. We build on this primitive and study \emph{agent-generated, CWE-conditioned} identifier bait across six autonomous agents and ten LLMs, with screened clean baselines, prompt-language variation, cross-model transfer, a prompt-level defense, and pre-completion detection; \cref{sec:rq8} shows that our bait is substantially more effective than XOXO on the same targets. Finally, unlike repository-level vulnerability detection~\cite{yildiz2025jitvul}, detecting \nametrap{} bait requires anticipating \emph{future} generation risk, because the distributed file contains no vulnerable statement.

\section{Methodology}\label{sec:method}

\subsection{Threat Model}

\textbf{Attacker Goal.} The attacker aims to make a victim AI coding agent generate code containing a targeted vulnerability while performing a legitimate development task. Rather than inserting vulnerabilities, the attacker renames identifiers in third-party Python code while preserving its functionality; the poisoned code then enters the victim agent's development context, where the modified identifiers influence subsequent code generation. We consider four target categories: OS command injection (CWE-78), cross-site scripting (CWE-79), use of a broken or risky cryptographic algorithm (CWE-327), and deserialization of untrusted data (CWE-502). An attack succeeds when the victim agent produces code containing the targeted vulnerability in response to an otherwise benign development request.

\textbf{Attacker Capabilities.} The attacker can prepare poisoned Python source code and publish it in a public GitHub repository, from which developers obtain it during legitimate development. The attacker may only rename identifiers and must preserve the original functionality. The attacker has no knowledge of the victim's prompt language, underlying LLM, or coding agent, and cannot control the victim's instructions or development environment. The poisoned source file is the attacker's only means of influence; whether and how it is used is determined entirely by the developer and the victim agent. Bait construction requires only black-box access to an attacker-chosen agent and model and at most ten renaming passes per file, so the attack scales cheaply to many files and projects.

\subsection{Dataset Collection and Preprocessing}
\label{sec:data_preprocessing}

We collect Python source files from the 100 most-starred GitHub projects. To identify suitable code locations, we analyze these files with Bandit and Semgrep for CWE-78, CWE-79, CWE-327, and CWE-502, and remove the lines associated with detected security risks, leaving incomplete code regions to be completed by AI coding agents. We select the first 50 files for each CWE, obtaining 200 source files from 31 GitHub projects that cover 47 project--CWE combinations. Before any identifier modification, we screen how coding agents complete the unmodified files. For each agent--model configuration, we run ten independent completion attempts with the same legitimate development prompt, restoring the preprocessed file and starting a fresh conversation for each attempt. The completed code is analyzed with Bandit and Semgrep, and a file is retained in the \emph{all-safe} set only if none of the ten attempts triggers the targeted CWE and the agent modifies the file in at least one attempt. The retained files constitute the clean baselines for all subsequent experiments. Data collection, code cleaning, and baseline screening are preparatory procedures; the attack begins only when the attacker-side agent modifies identifiers in a selected clean file.

\subsection{Agent-Driven Bait Construction}
\label{sec:agent_bait_construction}

\begin{figure*}[t]
    \centering
    \includegraphics[width=0.95\textwidth]{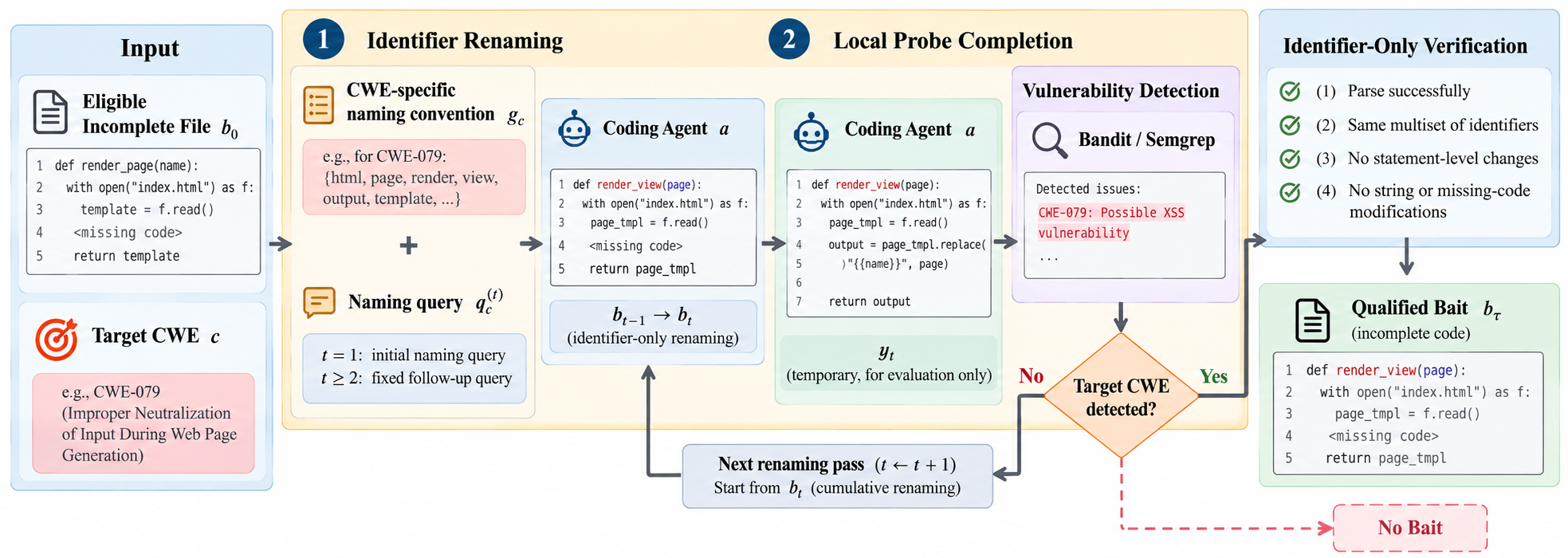}
    \caption{Overview of \nametrap{} bait construction. Given a preprocessed incomplete file and a target CWE, an attacker-side coding agent renames identifiers according to a CWE-specific naming convention (\ding{172}); a fresh agent session completes the renamed file in an isolated local probe, which is scanned with Bandit and Semgrep (\ding{173}). If the target CWE is not detected, another renaming pass follows. The first successful candidate that passes identifier-only verification is released as bait, i.e., the renamed source file \emph{before} completion.}
    \label{fig:baitcode_flow}
\end{figure*}

\paragraph{Attack Intuition.} Our attack exploits a mismatch between the execution semantics of source code and the contextual signals that AI coding agents extract from identifiers. In an incomplete source file, existing names implicitly suggest which operations, trust assumptions, and implementation conventions the missing code should follow. An attacker can therefore alter the apparent intent of the surrounding program through renaming alone, without inserting an instruction for the downstream agent or implementing a vulnerable operation. \Cref{fig:baitcode_flow} gives an overview of the bait-construction pipeline.

\paragraph{Constrained Attack Objective.} We formalize identifier-level poisoning as a constrained black-box optimization problem. For an eligible incomplete file $b_0$ and target CWE $c$, let $\mathcal{B}(b_0)$ denote the set of admissible identifier-renamed variants of $b_0$. A variant belongs to $\mathcal{B}(b_0)$ only if its renaming is consistent, its existing program behavior and incomplete regions are preserved, and no non-identifier content is altered. The attacker's ideal objective is
\begin{equation}
\label{eq:nametrap_objective}
\max_{b\in\mathcal{B}(b_0)}\;
\Pr_{(A_v,p)\sim\mathcal{V}}\!\left[
D_c\!\left(A_v(b,p)\right)=1
\right],
\end{equation}
where $A_v$ is a downstream coding agent, $p$ is a legitimate completion prompt, $D_c$ indicates whether the completed code is flagged for the targeted CWE, and $\mathcal{V}$ is the unknown distribution of victim agent--prompt configurations. Since the attacker observes neither the victim configuration nor the corresponding success probability, Equation~\eqref{eq:nametrap_objective} specifies the goal rather than a quantity we estimate directly. Instead, we conduct a bounded, agent-guided search that uses local probe feedback as a binary success signal.

\paragraph{CWE-Conditioned Naming Guidance.} Given an eligible file and a target CWE, we task a coding agent with aligning the file's identifiers to a predefined, CWE-specific naming convention, framing the operation as routine codebase maintenance performed by a senior engineer. Each convention introduces linguistic cues associated with an insecure implementation choice: command-execution and trusted-command terminology for CWE-78, prevalidated-output terminology for CWE-79, compatibility and fast-digest terminology for CWE-327, and trusted serialized-state terminology for CWE-502. The agent may rename functions, variables, and parameters, but must leave missing-code regions untouched and preserve all existing statements, string literals, comments, operators, line ordering, indentation, and control flow. Consequently, the payload that reaches the downstream agent consists only of modified identifiers, not the naming instruction.

\paragraph{Iterative Renaming and Local Completion Probes.} Bait construction proceeds through at most $T=10$ renaming passes. Let $\mathsf{Rename}_a$ denote the attacker-side agent's renaming operation and $q_c^{(t)}$ the CWE-conditioned naming query at pass $t$. The cumulative update is
\begin{equation}
\label{eq:nametrap_update}
b_t = \mathsf{Rename}_a\!\left(b_{t-1},q_c^{(t)}\right),\quad
q_c^{(t)} =
\begin{cases}
q_c^{\mathrm{init}}, & t=1,\\
q_c^{\mathrm{follow}}, & 2\le t\le T.
\end{cases}
\end{equation}
The first query requests naming alignment; the fixed follow-up query asks the agent to continue aligning the remaining identifiers. Each pass starts from $b_{t-1}$, not from the clean file $b_0$. After pass $t$, the incomplete candidate $b_t$ is copied into an isolated local probe, where a fresh coding-agent interaction completes the missing code using the standard legitimate prompt $p_{\mathrm{legit}}$:
\begin{equation}
\label{eq:nametrap_probe}
y_t = \mathsf{Complete}_{a}^{(t)}(b_t,p_{\mathrm{legit}}),\qquad
h_t = \mathbf{1}\!\left\{D_c(y_t)=1\right\},
\end{equation}
where $D_c(y_t)=1$ if Bandit or Semgrep reports the target CWE at medium severity or higher. The superscript $(t)$ denotes a fresh completion interaction, not an update to the completion model. Probing does not modify $b_t$ or the candidate used for the next pass; it supplies offline, black-box feedback about a candidate without any access to the eventual victim's model, agent framework, or prompt language.

\paragraph{Feedback-Guided Candidate Selection.} The search uses a first-success stopping rule rather than optimizing a differentiable loss or estimating the probability in Equation~\eqref{eq:nametrap_objective}. The first locally successful pass is
\begin{equation}
\label{eq:nametrap_stop}
\tau=\min\bigl\{t\in\{1,\ldots,T\}:h_t=1\bigr\},
\end{equation}
provided that the set is nonempty. The selected candidate is the \emph{pre-completion} file $b_{\tau}$, not the locally generated code $y_{\tau}$. If no probe triggers the target CWE within $T$ passes, the attempt returns no bait. Scanner feedback thus determines whether another renaming pass is needed and which candidate is retained.

\begin{table}[t]
\centering
\caption{Attack success rates (parseable ASR, \%) under four identifier-renaming conditions across six coding agents and four CWE categories.}
\label{tab:identifier_ablation}
\footnotesize
\setlength{\aboverulesep}{0pt}\setlength{\belowrulesep}{0pt}\setlength{\tabcolsep}{4pt}\renewcommand{\arraystretch}{1.0}
\begin{adjustbox}{max width=\columnwidth}
\begin{tabular}{llcccc>{\columncolor{allcol}}c}
\toprule
\rowcolor{hdrgray}\textbf{Agent} & \textbf{Variant} & \textbf{CWE-78} & \textbf{CWE-79} & \textbf{CWE-327} & \textbf{CWE-502} & \textbf{Overall} \\
\midrule
 & {\color{dimtext}Clean code} & {\color{dimtext}0} & {\color{dimtext}0} & {\color{dimtext}0} & {\color{dimtext}0} & {\color{dimtext}0} \\
 & Random renaming & 20.0 & 15.0 & 20.0 & 16.0 & 17.7 \\
 & Neutral renaming & 20.0 & 10.0 & 0.0 & 10.0 & 13.1 \\
\rowcolor{oursrow}\multirow{-4}{*}{\textbf{OpenCode}} & \textbf{\nametrap} & \textbf{32.0} & \textbf{30.0} & \textbf{60.0} & \textbf{24.0} & \textbf{30.8} \\
\midrule
 & {\color{dimtext}Clean code} & {\color{dimtext}0} & {\color{dimtext}0} & {\color{dimtext}0} & {\color{dimtext}0} & {\color{dimtext}0} \\
 & Random renaming & 1.7 & 0.0 & 60.0 & 5.0 & 10.0 \\
 & Neutral renaming & 1.7 & 0.0 & 20.0 & 10.0 & 5.0 \\
\rowcolor{oursrow}\multirow{-4}{*}{\textbf{Claude Code}} & \textbf{\nametrap} & \textbf{11.7} & \textbf{0.0} & \textbf{70.0} & \textbf{55.0} & \textbf{25.0} \\
\midrule
 & {\color{dimtext}Clean code} & {\color{dimtext}0} & {\color{dimtext}0} & {\color{dimtext}0} & {\color{dimtext}0} & {\color{dimtext}0} \\
 & Random renaming & 4.3 & 11.8 & 0.0 & 6.0 & 7.1 \\
 & Neutral renaming & 5.7 & 6.7 & 0.0 & 2.0 & 3.9 \\
\rowcolor{oursrow}\multirow{-4}{*}{\textbf{Codex}} & \textbf{\nametrap} & \textbf{15.7} & \textbf{37.5} & \textbf{0.0} & \textbf{42.0} & \textbf{24.7} \\
\midrule
 & {\color{dimtext}Clean code} & {\color{dimtext}0} & {\color{dimtext}0} & {\color{dimtext}0} & {\color{dimtext}0} & {\color{dimtext}0} \\
 & Random renaming & 0.0 & 2.5 & 10.0 & 10.0 & 5.6 \\
 & Neutral renaming & 0.0 & 2.5 & 33.3 & 0.0 & 8.0 \\
\rowcolor{oursrow}\multirow{-4}{*}{\textbf{mini-swe}} & \textbf{\nametrap} & \textbf{30.0} & \textbf{10.0} & \textbf{40.0} & \textbf{55.0} & \textbf{26.7} \\
\midrule
 & {\color{dimtext}Clean code} & {\color{dimtext}0} & {\color{dimtext}0} & {\color{dimtext}0} & {\color{dimtext}0} & {\color{dimtext}0} \\
 & Random renaming & 35.0 & 10.0 & 10.0 & 2.5 & 20.0 \\
 & Neutral renaming & 0.0 & 13.3 & 33.3 & 7.5 & 8.0 \\
\rowcolor{oursrow}\multirow{-4}{*}{\textbf{Aider}} & \textbf{\nametrap} & \textbf{100.0} & \textbf{63.3} & \textbf{60.0} & \textbf{27.5} & \textbf{56.0} \\
\midrule
 & {\color{dimtext}Clean code} & {\color{dimtext}0} & {\color{dimtext}0} & {\color{dimtext}0} & {\color{dimtext}0} & {\color{dimtext}0} \\
 & Random renaming & 10.0 & 5.0 & 17.6 & 0.0 & 10.4 \\
 & Neutral renaming & 0.0 & 0.0 & 5.6 & 0.0 & 1.3 \\
\rowcolor{oursrow}\multirow{-4}{*}{\textbf{OpenHands}} & \textbf{\nametrap} & \textbf{33.3} & \textbf{5.0} & \textbf{29.4} & \textbf{20.0} & \textbf{23.4} \\
\bottomrule
\end{tabular}
\end{adjustbox}
\end{table}
\begin{table}[t]
\centering
\caption{Attack success rates (ASR, \%) across six AI coding agents using \mdl{Qwen-3.5-27B-AWQ-4bit} on four CWE categories. G: gross ASR; P: parseable ASR. Bold: highest gross ASR per column.}
\label{tab:rq2}
\small
\setlength{\aboverulesep}{0pt}\setlength{\belowrulesep}{0pt}\setlength{\tabcolsep}{3pt}\renewcommand{\arraystretch}{1.05}
\begin{adjustbox}{max width=\columnwidth}
\begin{tabular}{l*{8}{c}>{\columncolor{allcol}}c>{\columncolor{allcol}}c}
\toprule
\rowcolor{hdrgray} & \multicolumn{2}{c}{\textbf{CWE-78}} & \multicolumn{2}{c}{\textbf{CWE-79}} & \multicolumn{2}{c}{\textbf{CWE-327}} & \multicolumn{2}{c}{\textbf{CWE-502}} & \multicolumn{2}{c}{\textbf{All}} \\
\rowcolor{hdrgray}\multirow{-2}{*}{\textbf{Agent}} & G & P & G & P & G & P & G & P & G & P \\
\midrule
OpenCode & 32.0 & {\color{dimtext}32.0} & 30.0 & {\color{dimtext}30.0} & 60.0 & {\color{dimtext}60.0} & 24.0 & {\color{dimtext}24.0} & 30.8 & {\color{dimtext}30.8} \\
Claude Code & 11.7 & {\color{dimtext}11.7} & 0.0 & {\color{dimtext}0.0} & \textbf{70.0} & {\color{dimtext}70.0} & \textbf{55.0} & {\color{dimtext}55.0} & 25.0 & {\color{dimtext}25.0} \\
Codex & 15.7 & {\color{dimtext}15.7} & 30.0 & {\color{dimtext}37.5} & 0.0 & {\color{dimtext}0.0} & 42.0 & {\color{dimtext}42.0} & 23.8 & {\color{dimtext}24.7} \\
mini-swe & 30.0 & {\color{dimtext}30.0} & 10.0 & {\color{dimtext}10.0} & 40.0 & {\color{dimtext}40.0} & \textbf{55.0} & {\color{dimtext}55.0} & 26.7 & {\color{dimtext}26.7} \\
Aider & \textbf{100.0} & {\color{dimtext}100.0} & \textbf{63.3} & {\color{dimtext}63.3} & 60.0 & {\color{dimtext}60.0} & 27.5 & {\color{dimtext}27.5} & \textbf{56.0} & {\color{dimtext}56.0} \\
OpenHands & 33.3 & {\color{dimtext}33.3} & 5.0 & {\color{dimtext}5.0} & 25.0 & {\color{dimtext}29.4} & 20.0 & {\color{dimtext}20.0} & 22.5 & {\color{dimtext}23.4} \\
\bottomrule
\end{tabular}
\end{adjustbox}
\end{table}
\begin{table*}[t]
\centering
\caption{Attack success rates (ASR, \%) across ten LLMs and three coding agents using English prompts. Bold: highest overall gross ASR per agent.}
\label{tab:rq4}
\footnotesize
\setlength{\aboverulesep}{0pt}\setlength{\belowrulesep}{0pt}\setlength{\tabcolsep}{4.5pt}\renewcommand{\arraystretch}{0.9}
\begin{adjustbox}{max width=\textwidth}
\begin{tabular}{cl*{8}{c}>{\columncolor{allcol}}c>{\columncolor{allcol}}c}
\toprule
\rowcolor{hdrgray} & & \multicolumn{2}{c}{\textbf{CWE-78}} & \multicolumn{2}{c}{\textbf{CWE-79}} & \multicolumn{2}{c}{\textbf{CWE-327}} & \multicolumn{2}{c}{\textbf{CWE-502}} & \multicolumn{2}{c}{\textbf{All}} \\
\rowcolor{hdrgray}\multirow{-2}{*}{\textbf{Agent}} & \multirow{-2}{*}{\textbf{Model}} & G & P & G & P & G & P & G & P & G & P \\
\midrule
 & \mdl{Qwen-3.5-27B-AWQ-4bit} & 32.0 & {\color{dimtext}32.0} & 30.0 & {\color{dimtext}30.0} & 60.0 & {\color{dimtext}60.0} & 24.0 & {\color{dimtext}24.0} & 30.8 & {\color{dimtext}30.8} \\
 & \mdl{Qwen-3.5-4B} & 20.0 & {\color{dimtext}40.0} & 10.0 & {\color{dimtext}10.0} & {\color{dimtext}---} & {\color{dimtext}---} & {\color{dimtext}---} & {\color{dimtext}---} & 15.0 & {\color{dimtext}20.0} \\
 & \mdl{Qwen-3.5-9B} & 0.0 & {\color{dimtext}0.0} & 10.0 & {\color{dimtext}10.0} & 50.0 & {\color{dimtext}50.0} & 50.0 & {\color{dimtext}58.8} & 32.0 & {\color{dimtext}34.8} \\
 & \mdl{Qwen-3.5-122B-A10B-AWQ-4bit} & 23.3 & {\color{dimtext}23.3} & 28.0 & {\color{dimtext}28.0} & {\color{dimtext}---} & {\color{dimtext}---} & 36.7 & {\color{dimtext}36.7} & 29.1 & {\color{dimtext}29.1} \\
 & \mdl{Gemma-4-26B-A4B} & 0.0 & {\color{dimtext}0.0} & {\color{dimtext}---} & {\color{dimtext}---} & 40.0 & {\color{dimtext}44.4} & 75.0 & {\color{dimtext}75.0} & 47.5 & {\color{dimtext}48.7} \\
 & \mdl{Gemma-4-31B} & 90.0 & {\color{dimtext}90.0} & {\color{dimtext}---} & {\color{dimtext}---} & 100.0 & {\color{dimtext}100.0} & 55.0 & {\color{dimtext}55.0} & \textbf{75.0} & {\color{dimtext}75.0} \\
 & \mdl{Nemotron-3-Nano-30B-A3B} & 10.0 & {\color{dimtext}10.0} & 0.0 & {\color{dimtext}0.0} & 20.0 & {\color{dimtext}20.7} & 16.7 & {\color{dimtext}16.7} & 12.7 & {\color{dimtext}12.8} \\
 & \mdl{GPT-5.4-mini} & 5.0 & {\color{dimtext}5.0} & {\color{dimtext}---} & {\color{dimtext}---} & 0.0 & {\color{dimtext}0.0} & 10.0 & {\color{dimtext}10.0} & 5.0 & {\color{dimtext}5.0} \\
 & \mdl{DeepSeek-V4-Flash-0731} & 7.5 & {\color{dimtext}7.5} & {\color{dimtext}---} & {\color{dimtext}---} & 30.0 & {\color{dimtext}30.0} & 20.0 & {\color{dimtext}20.0} & 11.0 & {\color{dimtext}11.0} \\
\multirow{-10}{*}{\rotatebox[origin=c]{90}{\textbf{OpenCode}}} & \mdl{Claude Haiku 4.5} & {\color{dimtext}---} & {\color{dimtext}---} & {\color{dimtext}---} & {\color{dimtext}---} & 40.0 & {\color{dimtext}40.0} & {\color{dimtext}---} & {\color{dimtext}---} & 40.0 & {\color{dimtext}40.0} \\
\midrule
 & \mdl{Qwen-3.5-27B-AWQ-4bit} & 11.7 & {\color{dimtext}11.7} & 0.0 & {\color{dimtext}0.0} & 70.0 & {\color{dimtext}70.0} & 55.0 & {\color{dimtext}55.0} & 25.0 & {\color{dimtext}25.0} \\
 & \mdl{Qwen-3.5-4B} & 0.0 & {\color{dimtext}0.0} & 6.7 & {\color{dimtext}7.1} & 10.0 & {\color{dimtext}10.0} & 33.3 & {\color{dimtext}37.0} & 16.4 & {\color{dimtext}17.7} \\
 & \mdl{Qwen-3.5-9B} & 30.0 & {\color{dimtext}31.3} & 6.0 & {\color{dimtext}6.2} & 100.0 & {\color{dimtext}100.0} & 17.5 & {\color{dimtext}21.9} & 24.1 & {\color{dimtext}26.1} \\
 & \mdl{Qwen-3.5-122B-A10B-AWQ-4bit} & 40.0 & {\color{dimtext}40.0} & 20.0 & {\color{dimtext}20.0} & {\color{dimtext}---} & {\color{dimtext}---} & {\color{dimtext}---} & {\color{dimtext}---} & 30.0 & {\color{dimtext}30.0} \\
 & \mdl{Gemma-4-26B-A4B} & 10.0 & {\color{dimtext}10.3} & 3.3 & {\color{dimtext}4.8} & 27.5 & {\color{dimtext}27.5} & 0.0 & {\color{dimtext}0.0} & 12.3 & {\color{dimtext}13.3} \\
 & \mdl{Gemma-4-31B} & 35.0 & {\color{dimtext}35.0} & 27.5 & {\color{dimtext}28.9} & 40.0 & {\color{dimtext}40.0} & 15.0 & {\color{dimtext}16.7} & 30.0 & {\color{dimtext}31.1} \\
 & \mdl{Nemotron-3-Nano-30B-A3B} & 25.0 & {\color{dimtext}30.3} & 0.0 & {\color{dimtext}0.0} & 10.0 & {\color{dimtext}10.3} & 55.0 & {\color{dimtext}78.6} & 24.0 & {\color{dimtext}29.6} \\
 & \mdl{GPT-5.4-mini} & 5.0 & {\color{dimtext}5.0} & 13.3 & {\color{dimtext}14.8} & 10.0 & {\color{dimtext}10.0} & 16.7 & {\color{dimtext}19.2} & 12.0 & {\color{dimtext}12.9} \\
 & \mdl{DeepSeek-V4-Flash-0731} & 15.0 & {\color{dimtext}16.7} & 10.0 & {\color{dimtext}11.1} & 10.0 & {\color{dimtext}10.0} & 10.0 & {\color{dimtext}11.1} & 12.2 & {\color{dimtext}13.1} \\
\multirow{-10}{*}{\rotatebox[origin=c]{90}{\textbf{Claude Code}}} & \mdl{Claude Haiku 4.5} & 20.0 & {\color{dimtext}20.0} & 16.7 & {\color{dimtext}16.7} & 5.0 & {\color{dimtext}5.9} & 100.0 & {\color{dimtext}100.0} & \textbf{33.3} & {\color{dimtext}34.5} \\
\midrule
 & \mdl{Qwen-3.5-27B-AWQ-4bit} & 15.7 & {\color{dimtext}15.7} & 30.0 & {\color{dimtext}37.5} & 0.0 & {\color{dimtext}0.0} & 42.0 & {\color{dimtext}42.0} & 23.8 & {\color{dimtext}24.7} \\
 & \mdl{Qwen-3.5-4B} & 4.3 & {\color{dimtext}4.3} & 13.3 & {\color{dimtext}14.3} & 12.5 & {\color{dimtext}18.5} & 22.0 & {\color{dimtext}23.9} & 12.1 & {\color{dimtext}13.5} \\
 & \mdl{Qwen-3.5-9B} & 16.7 & {\color{dimtext}16.7} & 18.0 & {\color{dimtext}18.3} & 6.7 & {\color{dimtext}7.4} & 3.3 & {\color{dimtext}3.3} & 13.7 & {\color{dimtext}14.0} \\
 & \mdl{Qwen-3.5-122B-A10B-AWQ-4bit} & 43.3 & {\color{dimtext}43.3} & 15.0 & {\color{dimtext}15.0} & 10.0 & {\color{dimtext}10.0} & 10.0 & {\color{dimtext}10.0} & 22.2 & {\color{dimtext}22.2} \\
 & \mdl{Gemma-4-26B-A4B} & 20.0 & {\color{dimtext}66.7} & 0.0 & {\color{dimtext}0.0} & 26.7 & {\color{dimtext}44.4} & {\color{dimtext}---} & {\color{dimtext}---} & 12.5 & {\color{dimtext}33.3} \\
 & \mdl{Gemma-4-31B} & 17.5 & {\color{dimtext}18.4} & 10.0 & {\color{dimtext}12.5} & 10.0 & {\color{dimtext}13.0} & 5.0 & {\color{dimtext}7.7} & 11.8 & {\color{dimtext}14.4} \\
 & \mdl{Nemotron-3-Nano-30B-A3B} & 5.0 & {\color{dimtext}50.0} & 5.0 & {\color{dimtext}20.0} & 10.0 & {\color{dimtext}50.0} & 17.5 & {\color{dimtext}77.8} & 11.0 & {\color{dimtext}55.0} \\
 & \mdl{GPT-5.4-mini} & 10.0 & {\color{dimtext}13.3} & 0.0 & {\color{dimtext}0.0} & 20.0 & {\color{dimtext}22.2} & 0.0 & {\color{dimtext}0.0} & 12.0 & {\color{dimtext}14.0} \\
 & \mdl{DeepSeek-V4-Flash-0731} & 28.3 & {\color{dimtext}28.3} & {\color{dimtext}---} & {\color{dimtext}---} & 0.0 & {\color{dimtext}0.0} & 30.0 & {\color{dimtext}30.0} & \textbf{25.0} & {\color{dimtext}25.0} \\
\multirow{-10}{*}{\rotatebox[origin=c]{90}{\textbf{Codex}}} & \mdl{Claude Haiku 4.5} & 5.0 & {\color{dimtext}5.0} & {\color{dimtext}---} & {\color{dimtext}---} & 5.0 & {\color{dimtext}5.3} & 12.5 & {\color{dimtext}12.5} & 8.8 & {\color{dimtext}8.9} \\
\bottomrule
\end{tabular}
\end{adjustbox}
\end{table*}
\paragraph{Identifier-Only Verification.} Candidate selection is followed by a strict admissibility check against the clean file $b_0$. Let $V(b,b_0)=1$ when token-level comparison, Git-based textual comparison, compilation checks, and manual adjudication all accept $b$ as an identifier-only, semantics-preserving modification, and $0$ otherwise. The final output is
\begin{equation}
\label{eq:nametrap_output}
b_{\mathrm{out}}=
\begin{cases}
b_{\tau}, & \tau\text{ exists and }V(b_{\tau},b_0)=1,\\
\varnothing, & \text{otherwise}.
\end{cases}
\end{equation}
Only qualified candidates are retained. The output is always the incomplete, identifier-renamed source file, never the vulnerable completion $y_{\tau}$; the targeted vulnerability emerges only when a downstream coding agent later completes the file.

\section{Experimental Evaluation}
\label{sec:evaluation}

\subsection{Experimental Settings}
\label{sec:experimental_settings}

\textbf{Coding Agents and Language Models.}
We evaluate six AI coding agents: OpenCode, Claude Code, Codex, mini-swe-agent (mini-swe), Aider, and OpenHands. To study the influence of the underlying model, we further evaluate ten LLMs with three agent frameworks (OpenCode, Claude Code, and Codex): \mdl{Qwen-3.5-4B}, \mdl{Qwen-3.5-9B}, \mdl{Qwen-3.5-27B-AWQ-4bit}, \mdl{Qwen-3.5-122B-A10B-AWQ-4bit}, \mdl{Gemma-4-26B-A4B}, \mdl{Gemma-4-31B}, \mdl{Nemotron-3-Nano-30B-A3B}, \mdl{GPT-5.4-mini}, \mdl{DeepSeek-V4-Flash-0731}, and \mdl{Claude Haiku 4.5}. Unless stated otherwise, the six-agent experiments use \mdl{Qwen-3.5-27B-AWQ-4bit}.

\textbf{Victim Evaluation and Metrics.}
Each qualified bait file is completed by the victim agent with the legitimate development prompt, starting a fresh session for every attempt. Each completed file is scanned with Bandit and Semgrep, and an attempt succeeds if the target CWE is reported at medium severity or higher. We report the \textbf{gross ASR (G)}, computed over all completion attempts, and the \textbf{parseable ASR (P)}, computed over attempts whose output can be analyzed by the scanners. \emph{Overall}/\emph{All} pools all attempts of a configuration across CWEs, and a dash (---) indicates that no qualifying bait was obtained for that configuration and CWE. Unless stated otherwise, bait is constructed with the same agent--model configuration that is evaluated as the victim; RQ5 decouples the two.

\textbf{Detection Tools and Hardware.}
To assess whether poisoned files can be flagged before use, we evaluate three AI-based and four static analyzers (\cref{sec:rq7}). All experiments were conducted on servers equipped with NVIDIA H100 and H200 GPUs.

\begin{table}[t]
\centering
\caption{Attack success rates (ASR, \%) across six AI coding agents using \mdl{Qwen-3.5-27B-AWQ-4bit} with Traditional Chinese prompts. Bold: highest gross ASR per column.}
\label{tab:rq3}
\small
\setlength{\aboverulesep}{0pt}\setlength{\belowrulesep}{0pt}\setlength{\tabcolsep}{3pt}\renewcommand{\arraystretch}{1.05}
\begin{adjustbox}{max width=\columnwidth}
\begin{tabular}{l*{8}{c}>{\columncolor{allcol}}c>{\columncolor{allcol}}c}
\toprule
\rowcolor{hdrgray} & \multicolumn{2}{c}{\textbf{CWE-78}} & \multicolumn{2}{c}{\textbf{CWE-79}} & \multicolumn{2}{c}{\textbf{CWE-327}} & \multicolumn{2}{c}{\textbf{CWE-502}} & \multicolumn{2}{c}{\textbf{All}} \\
\rowcolor{hdrgray}\multirow{-2}{*}{\textbf{Agent}} & G & P & G & P & G & P & G & P & G & P \\
\midrule
OpenCode & 46.0 & {\color{dimtext}46.0} & 20.0 & {\color{dimtext}20.0} & \textbf{70.0} & {\color{dimtext}70.0} & 28.0 & {\color{dimtext}28.0} & 36.9 & {\color{dimtext}36.9} \\
Claude Code & 20.0 & {\color{dimtext}20.0} & 0.0 & {\color{dimtext}0.0} & \textbf{70.0} & {\color{dimtext}70.0} & 50.0 & {\color{dimtext}50.0} & 29.0 & {\color{dimtext}29.0} \\
Codex & 14.3 & {\color{dimtext}14.3} & 25.0 & {\color{dimtext}29.4} & 5.0 & {\color{dimtext}5.3} & 40.0 & {\color{dimtext}40.0} & 22.5 & {\color{dimtext}23.1} \\
mini-swe & 20.0 & {\color{dimtext}20.0} & 10.0 & {\color{dimtext}10.0} & 40.0 & {\color{dimtext}40.0} & \textbf{55.0} & {\color{dimtext}55.0} & 26.7 & {\color{dimtext}26.7} \\
Aider & \textbf{50.0} & {\color{dimtext}50.0} & \textbf{30.0} & {\color{dimtext}30.0} & 40.0 & {\color{dimtext}40.0} & 35.0 & {\color{dimtext}35.9} & \textbf{37.0} & {\color{dimtext}37.4} \\
OpenHands & 26.7 & {\color{dimtext}26.7} & 10.0 & {\color{dimtext}10.0} & 15.0 & {\color{dimtext}15.0} & 30.0 & {\color{dimtext}30.0} & 20.0 & {\color{dimtext}20.0} \\
\bottomrule
\end{tabular}
\end{adjustbox}
\end{table}
\begin{table*}[t]
\centering
\caption{Cross-model transferability of bait code on CWE-78 (OS command injection). Rows: model used to construct the bait; columns: model that completes the task. Each cell reports gross/parseable ASR (\%); shading scales with gross ASR, and bold marks the source model.}
\label{tab:cross_models_cwe78}
\scriptsize
\setlength{\aboverulesep}{0pt}\setlength{\belowrulesep}{0pt}\setlength{\tabcolsep}{3pt}\renewcommand{\arraystretch}{0.92}
\begin{adjustbox}{max width=\textwidth}
\begin{tabular}{cl*{10}{c}}
\toprule
\rowcolor{hdrgray}\rule{0pt}{2.4ex} & & \multicolumn{10}{c}{\textbf{Task-completion model}} \\
\cmidrule(lr){3-12}
\rowcolor{hdrgray} & & \texttt{Qwen-3.5-} & \texttt{Qwen-3.5-} & \texttt{Qwen-3.5-} & \texttt{Qwen-3.5-122B-} & \texttt{Gemma-4-} & \texttt{Gemma-4-} & \texttt{Nemotron-3-} & \texttt{GPT-5.4-} & \texttt{DeepSeek-V4-} & \texttt{Claude} \\
\rowcolor{hdrgray}\multirow{-3}{*}{\textbf{Agent}} & \multirow{-3}{*}{\textbf{Bait model}} & \texttt{27B-AWQ-4bit} & \texttt{4B} & \texttt{9B} & \texttt{A10B-AWQ-4bit} & \texttt{26B-A4B} & \texttt{31B} & \texttt{Nano-30B-A3B} & \texttt{mini} & \texttt{Flash-0731} & \texttt{Haiku 4.5} \\
\midrule
 & \mdl{Qwen-3.5-27B-AWQ-4bit} & \cellcolor{heat!13}\textbf{32.0} / 32.0 & \cellcolor{heat!6}14.0 / 14.6 & \cellcolor{heat!8}20.0 / 20.4 & \cellcolor{heat!4}4.0 / 4.0 & 0.0 / 0.0 & \cellcolor{heat!6}14.0 / 14.3 & 0.0 / 0.0 & 0.0 / 0.0 & \cellcolor{heat!21}50.0 / 50.0 & \cellcolor{heat!6}14.0 / 14.0 \\
 & \mdl{Qwen-3.5-4B} & \cellcolor{heat!4}10.0 / 10.0 & \cellcolor{heat!8}\textbf{20.0} / 40.0 & 0.0 / 0.0 & \cellcolor{heat!42}100.0 / 100.0 & 0.0 / 0.0 & 0.0 / 0.0 & 0.0 / 0.0 & 0.0 / 0.0 & 0.0 / 0.0 & 0.0 / 0.0 \\
 & \mdl{Qwen-3.5-9B} & 0.0 / 0.0 & \cellcolor{heat!4}10.0 / 10.0 & \textbf{0.0} / 0.0 & 0.0 / 0.0 & 0.0 / 0.0 & 0.0 / 0.0 & 0.0 / 0.0 & 0.0 / 0.0 & 0.0 / 0.0 & 0.0 / 0.0 \\
 & \mdl{Qwen-3.5-122B-A10B-AWQ-4bit} & \cellcolor{heat!4}6.7 / 6.7 & \cellcolor{heat!4}10.0 / 10.0 & \cellcolor{heat!4}10.0 / 10.3 & \cellcolor{heat!10}\textbf{23.3} / 23.3 & \cellcolor{heat!11}26.7 / 26.7 & \cellcolor{heat!6}13.3 / 13.3 & 0.0 / 0.0 & 0.0 / 0.0 & \cellcolor{heat!6}13.3 / 13.3 & \cellcolor{heat!4}3.3 / 3.3 \\
 & \mdl{Gemma-4-26B-A4B} & 0.0 / 0.0 & 0.0 / 0.0 & 0.0 / 0.0 & 0.0 / 0.0 & \textbf{0.0} / 0.0 & 0.0 / 0.0 & 0.0 / 0.0 & 0.0 / 0.0 & 0.0 / 0.0 & 0.0 / 0.0 \\
 & \mdl{Gemma-4-31B} & 0.0 / 0.0 & \cellcolor{heat!4}10.0 / 10.0 & \cellcolor{heat!8}20.0 / 22.2 & 0.0 / 0.0 & 0.0 / 0.0 & \cellcolor{heat!38}\textbf{90.0} / 90.0 & 0.0 / 0.0 & 0.0 / 0.0 & 0.0 / 0.0 & 0.0 / 0.0 \\
 & \mdl{Nemotron-3-Nano-30B-A3B} & \cellcolor{heat!18}43.3 / 43.3 & 0.0 / 0.0 & \cellcolor{heat!14}33.3 / 33.3 & \cellcolor{heat!10}23.3 / 23.3 & \cellcolor{heat!10}23.3 / 23.3 & \cellcolor{heat!22}53.3 / 53.3 & \cellcolor{heat!4}\textbf{10.0} / 10.0 & \cellcolor{heat!4}10.0 / 10.0 & \cellcolor{heat!6}13.3 / 13.3 & \cellcolor{heat!13}30.0 / 30.0 \\
 & \mdl{GPT-5.4-mini} & \cellcolor{heat!23}55.0 / 55.0 & \cellcolor{heat!4}5.0 / 5.0 & \cellcolor{heat!4}10.0 / 10.0 & \cellcolor{heat!15}35.0 / 35.0 & 0.0 / 0.0 & \cellcolor{heat!19}45.0 / 45.0 & \cellcolor{heat!4}5.0 / 5.0 & \cellcolor{heat!4}\textbf{5.0} / 5.0 & \cellcolor{heat!13}30.0 / 30.0 & \cellcolor{heat!19}45.0 / 45.0 \\
 & \mdl{DeepSeek-V4-Flash-0731} & \cellcolor{heat!6}15.0 / 15.0 & \cellcolor{heat!7}16.2 / 16.2 & \cellcolor{heat!4}5.0 / 5.1 & \cellcolor{heat!4}5.0 / 5.0 & \cellcolor{heat!4}6.2 / 6.2 & \cellcolor{heat!4}6.2 / 6.2 & \cellcolor{heat!4}1.2 / 1.5 & 0.0 / 0.0 & \cellcolor{heat!4}\textbf{7.5} / 7.5 & \cellcolor{heat!4}5.0 / 5.0 \\
\multirow{-10}{*}{\rotatebox[origin=c]{90}{\textbf{OpenCode}}} & \mdl{Claude Haiku 4.5} & {\color{dimtext}---} & {\color{dimtext}---} & {\color{dimtext}---} & {\color{dimtext}---} & {\color{dimtext}---} & {\color{dimtext}---} & {\color{dimtext}---} & {\color{dimtext}---} & {\color{dimtext}---} & {\color{dimtext}---} \\
\midrule
 & \mdl{Qwen-3.5-27B-AWQ-4bit} & \cellcolor{heat!5}\textbf{11.7} / 11.7 & \cellcolor{heat!7}16.7 / 16.7 & \cellcolor{heat!7}16.7 / 16.9 & \cellcolor{heat!6}13.3 / 13.3 & \cellcolor{heat!6}15.0 / 15.0 & \cellcolor{heat!7}16.7 / 18.5 & \cellcolor{heat!4}6.7 / 6.9 & \cellcolor{heat!4}5.0 / 5.1 & \cellcolor{heat!8}20.0 / 20.3 & \cellcolor{heat!4}8.3 / 8.3 \\
 & \mdl{Qwen-3.5-4B} & 0.0 / 0.0 & \textbf{0.0} / 0.0 & \cellcolor{heat!4}10.0 / 10.0 & 0.0 / 0.0 & 0.0 / 0.0 & 0.0 / 0.0 & \cellcolor{heat!4}10.0 / 10.8 & \cellcolor{heat!4}2.5 / 2.5 & \cellcolor{heat!4}10.0 / 10.5 & 0.0 / 0.0 \\
 & \mdl{Qwen-3.5-9B} & \cellcolor{heat!4}2.9 / 2.9 & \cellcolor{heat!8}18.6 / 20.3 & \cellcolor{heat!13}\textbf{30.0} / 31.3 & \cellcolor{heat!4}5.7 / 5.7 & \cellcolor{heat!4}5.7 / 6.2 & 0.0 / 0.0 & \cellcolor{heat!5}12.9 / 14.1 & \cellcolor{heat!4}1.4 / 1.4 & \cellcolor{heat!4}2.9 / 2.9 & 0.0 / 0.0 \\
 & \mdl{Qwen-3.5-122B-A10B-AWQ-4bit} & \cellcolor{heat!10}25.0 / 25.0 & \cellcolor{heat!4}10.0 / 11.1 & \cellcolor{heat!6}15.0 / 15.0 & \cellcolor{heat!17}\textbf{40.0} / 40.0 & \cellcolor{heat!4}5.0 / 5.0 & \cellcolor{heat!10}25.0 / 25.0 & \cellcolor{heat!8}20.0 / 26.7 & 0.0 / 0.0 & \cellcolor{heat!4}5.0 / 5.3 & \cellcolor{heat!6}15.0 / 15.0 \\
 & \mdl{Gemma-4-26B-A4B} & \cellcolor{heat!14}32.5 / 32.5 & \cellcolor{heat!4}7.5 / 8.3 & \cellcolor{heat!5}12.5 / 12.8 & 0.0 / 0.0 & \cellcolor{heat!4}\textbf{10.0} / 10.3 & 0.0 / 0.0 & 0.0 / 0.0 & 0.0 / 0.0 & \cellcolor{heat!19}45.0 / 46.2 & \cellcolor{heat!17}40.0 / 40.0 \\
 & \mdl{Gemma-4-31B} & 0.0 / 0.0 & \cellcolor{heat!4}5.0 / 5.0 & \cellcolor{heat!4}5.0 / 5.0 & 0.0 / 0.0 & 0.0 / 0.0 & \cellcolor{heat!15}\textbf{35.0} / 35.0 & 0.0 / 0.0 & 0.0 / 0.0 & 0.0 / 0.0 & 0.0 / 0.0 \\
 & \mdl{Nemotron-3-Nano-30B-A3B} & \cellcolor{heat!4}10.0 / 10.0 & \cellcolor{heat!8}20.0 / 20.5 & \cellcolor{heat!5}12.5 / 12.5 & \cellcolor{heat!6}15.0 / 15.0 & \cellcolor{heat!9}22.5 / 28.1 & \cellcolor{heat!4}10.0 / 18.2 & \cellcolor{heat!10}\textbf{25.0} / 30.3 & \cellcolor{heat!6}15.0 / 15.0 & \cellcolor{heat!12}27.5 / 30.6 & \cellcolor{heat!7}17.5 / 17.5 \\
 & \mdl{GPT-5.4-mini} & \cellcolor{heat!21}50.0 / 50.0 & \cellcolor{heat!8}20.0 / 20.0 & \cellcolor{heat!6}15.0 / 15.0 & 0.0 / 0.0 & 0.0 / 0.0 & \cellcolor{heat!4}5.0 / 5.0 & \cellcolor{heat!4}10.0 / 13.3 & \cellcolor{heat!4}\textbf{5.0} / 5.0 & \cellcolor{heat!15}35.0 / 41.2 & \cellcolor{heat!19}45.0 / 45.0 \\
 & \mdl{DeepSeek-V4-Flash-0731} & \cellcolor{heat!5}12.5 / 12.5 & \cellcolor{heat!8}20.0 / 20.0 & \cellcolor{heat!8}20.0 / 20.5 & \cellcolor{heat!4}5.0 / 5.0 & \cellcolor{heat!6}15.0 / 18.2 & \cellcolor{heat!6}15.0 / 18.8 & \cellcolor{heat!7}17.5 / 17.5 & \cellcolor{heat!5}12.5 / 12.5 & \cellcolor{heat!6}\textbf{15.0} / 16.7 & \cellcolor{heat!7}17.5 / 17.5 \\
\multirow{-10}{*}{\rotatebox[origin=c]{90}{\textbf{Claude Code}}} & \mdl{Claude Haiku 4.5} & \cellcolor{heat!15}35.0 / 35.0 & \cellcolor{heat!4}10.0 / 10.0 & \cellcolor{heat!6}15.0 / 15.8 & \cellcolor{heat!6}15.0 / 15.0 & 0.0 / 0.0 & \cellcolor{heat!6}15.0 / 23.1 & 0.0 / 0.0 & 0.0 / 0.0 & \cellcolor{heat!13}30.0 / 30.0 & \cellcolor{heat!8}\textbf{20.0} / 20.0 \\
\midrule
 & \mdl{Qwen-3.5-27B-AWQ-4bit} & \cellcolor{heat!7}\textbf{15.7} / 15.7 & \cellcolor{heat!4}10.0 / 11.3 & \cellcolor{heat!10}24.3 / 24.3 & \cellcolor{heat!4}1.4 / 1.4 & \cellcolor{heat!4}5.7 / 12.9 & \cellcolor{heat!4}4.3 / 5.3 & \cellcolor{heat!4}1.4 / 9.1 & \cellcolor{heat!4}1.4 / 1.4 & \cellcolor{heat!4}7.1 / 7.1 & \cellcolor{heat!4}2.9 / 2.9 \\
 & \mdl{Qwen-3.5-4B} & \cellcolor{heat!4}4.3 / 4.3 & \cellcolor{heat!4}\textbf{4.3} / 4.3 & \cellcolor{heat!5}12.9 / 12.9 & 0.0 / 0.0 & 0.0 / 0.0 & 0.0 / 0.0 & 0.0 / 0.0 & 0.0 / 0.0 & 0.0 / 0.0 & 0.0 / 0.0 \\
 & \mdl{Qwen-3.5-9B} & \cellcolor{heat!4}8.9 / 8.9 & \cellcolor{heat!4}10.1 / 11.0 & \cellcolor{heat!7}\textbf{16.7} / 16.7 & \cellcolor{heat!4}1.1 / 1.1 & \cellcolor{heat!4}1.1 / 1.9 & \cellcolor{heat!4}3.3 / 3.5 & \cellcolor{heat!4}1.1 / 6.7 & \cellcolor{heat!4}1.1 / 1.1 & \cellcolor{heat!6}14.4 / 14.4 & \cellcolor{heat!4}6.7 / 6.7 \\
 & \mdl{Qwen-3.5-122B-A10B-AWQ-4bit} & \cellcolor{heat!11}26.7 / 26.7 & \cellcolor{heat!4}3.3 / 4.3 & \cellcolor{heat!11}26.7 / 29.6 & \cellcolor{heat!18}\textbf{43.3} / 43.3 & \cellcolor{heat!4}6.7 / 14.3 & \cellcolor{heat!13}30.0 / 31.0 & 0.0 / 0.0 & 0.0 / 0.0 & \cellcolor{heat!4}6.7 / 6.7 & \cellcolor{heat!4}6.7 / 6.7 \\
 & \mdl{Gemma-4-26B-A4B} & 0.0 / 0.0 & 0.0 / 0.0 & \cellcolor{heat!8}20.0 / 20.0 & 0.0 / 0.0 & \cellcolor{heat!8}\textbf{20.0} / 66.7 & \cellcolor{heat!8}20.0 / 25.0 & 0.0 / 0.0 & 0.0 / 0.0 & 0.0 / 0.0 & 0.0 / 0.0 \\
 & \mdl{Gemma-4-31B} & \cellcolor{heat!12}27.5 / 27.5 & \cellcolor{heat!4}2.5 / 2.9 & \cellcolor{heat!4}5.0 / 5.0 & \cellcolor{heat!4}7.5 / 7.5 & 0.0 / 0.0 & \cellcolor{heat!7}\textbf{17.5} / 18.4 & 0.0 / 0.0 & \cellcolor{heat!4}7.5 / 8.6 & \cellcolor{heat!22}52.5 / 53.8 & \cellcolor{heat!19}45.0 / 45.0 \\
 & \mdl{Nemotron-3-Nano-30B-A3B} & 0.0 / 0.0 & \cellcolor{heat!4}10.0 / 13.3 & \cellcolor{heat!8}20.0 / 20.0 & 0.0 / 0.0 & 0.0 / 0.0 & \cellcolor{heat!6}15.0 / 16.7 & \cellcolor{heat!4}\textbf{5.0} / 50.0 & 0.0 / 0.0 & \cellcolor{heat!13}30.0 / 30.0 & 0.0 / 0.0 \\
 & \mdl{GPT-5.4-mini} & \cellcolor{heat!15}35.0 / 35.0 & 0.0 / 0.0 & 0.0 / 0.0 & \cellcolor{heat!4}10.0 / 10.0 & \cellcolor{heat!4}5.0 / 7.7 & \cellcolor{heat!19}45.0 / 45.0 & 0.0 / 0.0 & \cellcolor{heat!4}\textbf{10.0} / 13.3 & \cellcolor{heat!36}85.0 / 85.0 & \cellcolor{heat!27}65.0 / 65.0 \\
 & \mdl{DeepSeek-V4-Flash-0731} & \cellcolor{heat!6}15.0 / 15.0 & \cellcolor{heat!5}11.7 / 13.0 & \cellcolor{heat!8}18.3 / 18.6 & \cellcolor{heat!5}11.7 / 11.7 & \cellcolor{heat!4}3.3 / 10.0 & \cellcolor{heat!6}15.0 / 18.4 & 0.0 / 0.0 & \cellcolor{heat!4}6.7 / 7.0 & \cellcolor{heat!12}\textbf{28.3} / 28.3 & \cellcolor{heat!4}6.7 / 6.7 \\
\multirow{-10}{*}{\rotatebox[origin=c]{90}{\textbf{Codex}}} & \mdl{Claude Haiku 4.5} & \cellcolor{heat!13}30.0 / 30.0 & \cellcolor{heat!13}30.0 / 31.6 & \cellcolor{heat!23}55.0 / 55.0 & \cellcolor{heat!8}20.0 / 20.0 & \cellcolor{heat!6}15.0 / 42.9 & \cellcolor{heat!19}45.0 / 50.0 & \cellcolor{heat!6}15.0 / 60.0 & \cellcolor{heat!8}20.0 / 20.0 & \cellcolor{heat!17}40.0 / 40.0 & \cellcolor{heat!4}\textbf{5.0} / 5.0 \\
\bottomrule
\end{tabular}
\end{adjustbox}
\end{table*}
\subsection{RQ1: Impact of Identifier Renaming Strategies}\label{sec:rq1}

We first test whether attack success stems from deliberately crafted names or from the mere act of renaming. We compare four variants of each file: Clean Code, Random Renaming (randomly generated names), Neutral Renaming (semantically neutral names), and Ours (agent-generated, CWE-guided names), across six coding agents and four CWE categories.

As shown in \cref{tab:identifier_ablation}, our method achieves higher overall ASRs than both random and neutral renaming for all six agents. The largest gain occurs with Aider, where our method reaches 56.0\% versus 20.0\% for random and 8.0\% for neutral renaming; for Codex, the overall ASR rises from 7.1\% and 3.9\% to 24.7\%. Uninformed renaming is not entirely benign: random renaming reaches 60.0\% on CWE-327 with Claude Code and 35.0\% on CWE-78 with Aider, indicating that disrupting the original names alone can perturb agent behavior. Nevertheless, CWE-guided names match or exceed both baselines for every agent--CWE pair, showing that the attack's effectiveness stems from adversarial name selection rather than from renaming per se. Averaged over the six agents, the overall ASR is 31.1\% for our method, compared with 11.8\% for random and 6.6\% for neutral renaming.

\subsection{RQ2: Impact of Coding Agents}
\label{sec:rq2}

To investigate how the agent framework affects attack effectiveness, we evaluate six agents with the same underlying model, \mdl{Qwen-3.5-27B-AWQ-4bit}, across four CWE categories. As shown in \cref{tab:rq2}, effectiveness varies substantially across agents. Aider achieves the highest overall gross ASR of 56.0\%, while the remaining agents range from 22.5\% to 30.8\%. Every agent yields qualifying bait for all four CWEs in this setting. CWE-79 is the most challenging target: Claude Code attains 0.0\%, OpenHands 5.0\%, and mini-swe 10.0\%, although Aider still reaches 63.3\%. CWE-327 ranges from 0.0\% with Codex to 70.0\% with Claude Code. In contrast, CWE-78 attains nonzero ASRs for all six agents, with Aider reaching 100.0\%. Attack effectiveness thus depends on both the agent framework and the targeted CWE. Gross and parseable ASRs coincide except for Codex on CWE-79 (30.0\% vs.\ 37.5\%) and OpenHands on CWE-327 (25.0\% vs.\ 29.4\%), so nearly all completions in this setting are analyzable and the reported ASRs are not inflated by unparseable outputs.

\begin{table*}[t]
\centering
\caption{Attack success rates (ASR, \%) across ten LLMs and three coding agents with a prompt-based defense. Bold: highest overall gross ASR per agent.}
\label{tab:rq6-1}
\footnotesize
\setlength{\aboverulesep}{0pt}\setlength{\belowrulesep}{0pt}\setlength{\tabcolsep}{4.5pt}\renewcommand{\arraystretch}{0.9}
\begin{adjustbox}{max width=\textwidth}
\begin{tabular}{cl*{8}{c}>{\columncolor{allcol}}c>{\columncolor{allcol}}c}
\toprule
\rowcolor{hdrgray} & & \multicolumn{2}{c}{\textbf{CWE-78}} & \multicolumn{2}{c}{\textbf{CWE-79}} & \multicolumn{2}{c}{\textbf{CWE-327}} & \multicolumn{2}{c}{\textbf{CWE-502}} & \multicolumn{2}{c}{\textbf{All}} \\
\rowcolor{hdrgray}\multirow{-2}{*}{\textbf{Agent}} & \multirow{-2}{*}{\textbf{Model}} & G & P & G & P & G & P & G & P & G & P \\
\midrule
 & \mdl{Qwen-3.5-27B-AWQ-4bit} & 32.0 & {\color{dimtext}32.0} & 15.0 & {\color{dimtext}15.0} & 0.0 & {\color{dimtext}0.0} & 16.0 & {\color{dimtext}16.0} & 20.8 & {\color{dimtext}20.8} \\
 & \mdl{Qwen-3.5-4B} & 40.0 & {\color{dimtext}44.4} & 0.0 & {\color{dimtext}0.0} & {\color{dimtext}---} & {\color{dimtext}---} & {\color{dimtext}---} & {\color{dimtext}---} & 20.0 & {\color{dimtext}21.1} \\
 & \mdl{Qwen-3.5-9B} & 20.0 & {\color{dimtext}20.0} & 30.0 & {\color{dimtext}30.0} & 30.0 & {\color{dimtext}30.0} & 45.0 & {\color{dimtext}45.0} & 34.0 & {\color{dimtext}34.0} \\
 & \mdl{Qwen-3.5-122B-A10B-AWQ-4bit} & 10.0 & {\color{dimtext}10.0} & 26.0 & {\color{dimtext}26.0} & {\color{dimtext}---} & {\color{dimtext}---} & 23.3 & {\color{dimtext}23.3} & 20.9 & {\color{dimtext}20.9} \\
 & \mdl{Gemma-4-26B-A4B} & 0.0 & {\color{dimtext}0.0} & {\color{dimtext}---} & {\color{dimtext}---} & 80.0 & {\color{dimtext}88.9} & 85.0 & {\color{dimtext}85.0} & \textbf{62.5} & {\color{dimtext}64.1} \\
 & \mdl{Gemma-4-31B} & 20.0 & {\color{dimtext}20.0} & {\color{dimtext}---} & {\color{dimtext}---} & 70.0 & {\color{dimtext}70.0} & 15.0 & {\color{dimtext}15.0} & 30.0 & {\color{dimtext}30.0} \\
 & \mdl{Nemotron-3-Nano-30B-A3B} & 10.0 & {\color{dimtext}11.1} & 0.0 & {\color{dimtext}0.0} & 10.0 & {\color{dimtext}10.7} & 23.3 & {\color{dimtext}23.3} & 11.8 & {\color{dimtext}12.4} \\
 & \mdl{GPT-5.4-mini} & 0.0 & {\color{dimtext}0.0} & {\color{dimtext}---} & {\color{dimtext}---} & 0.0 & {\color{dimtext}0.0} & 0.0 & {\color{dimtext}0.0} & 0.0 & {\color{dimtext}0.0} \\
 & \mdl{DeepSeek-V4-Flash-0731} & 3.8 & {\color{dimtext}3.8} & {\color{dimtext}---} & {\color{dimtext}---} & 20.0 & {\color{dimtext}20.0} & 40.0 & {\color{dimtext}40.0} & 9.0 & {\color{dimtext}9.0} \\
\multirow{-10}{*}{\rotatebox[origin=c]{90}{\textbf{OpenCode}}} & \mdl{Claude Haiku 4.5} & {\color{dimtext}---} & {\color{dimtext}---} & {\color{dimtext}---} & {\color{dimtext}---} & 0.0 & {\color{dimtext}0.0} & {\color{dimtext}---} & {\color{dimtext}---} & 0.0 & {\color{dimtext}0.0} \\
\midrule
 & \mdl{Qwen-3.5-27B-AWQ-4bit} & 20.0 & {\color{dimtext}20.0} & 10.0 & {\color{dimtext}10.0} & 70.0 & {\color{dimtext}70.0} & 45.0 & {\color{dimtext}45.0} & 29.0 & {\color{dimtext}29.0} \\
 & \mdl{Qwen-3.5-4B} & 2.5 & {\color{dimtext}2.6} & 10.0 & {\color{dimtext}10.3} & 10.0 & {\color{dimtext}10.0} & 31.7 & {\color{dimtext}32.8} & 17.1 & {\color{dimtext}17.8} \\
 & \mdl{Qwen-3.5-9B} & 18.6 & {\color{dimtext}18.8} & 10.0 & {\color{dimtext}10.2} & 80.0 & {\color{dimtext}80.0} & 25.0 & {\color{dimtext}32.3} & 21.2 & {\color{dimtext}22.6} \\
 & \mdl{Qwen-3.5-122B-A10B-AWQ-4bit} & 50.0 & {\color{dimtext}50.0} & 25.0 & {\color{dimtext}25.0} & {\color{dimtext}---} & {\color{dimtext}---} & {\color{dimtext}---} & {\color{dimtext}---} & \textbf{37.5} & {\color{dimtext}37.5} \\
 & \mdl{Gemma-4-26B-A4B} & 5.0 & {\color{dimtext}5.0} & 3.3 & {\color{dimtext}4.3} & 22.5 & {\color{dimtext}23.7} & 20.0 & {\color{dimtext}20.0} & 12.3 & {\color{dimtext}13.2} \\
 & \mdl{Gemma-4-31B} & 0.0 & {\color{dimtext}0.0} & 25.0 & {\color{dimtext}27.0} & 6.7 & {\color{dimtext}6.7} & 25.0 & {\color{dimtext}27.8} & 15.5 & {\color{dimtext}16.2} \\
 & \mdl{Nemotron-3-Nano-30B-A3B} & 22.5 & {\color{dimtext}31.0} & 0.0 & {\color{dimtext}0.0} & 16.7 & {\color{dimtext}17.9} & 55.0 & {\color{dimtext}84.6} & 25.0 & {\color{dimtext}33.8} \\
 & \mdl{GPT-5.4-mini} & 10.0 & {\color{dimtext}10.0} & 16.7 & {\color{dimtext}19.2} & 0.0 & {\color{dimtext}0.0} & 13.3 & {\color{dimtext}15.4} & 11.0 & {\color{dimtext}12.0} \\
 & \mdl{DeepSeek-V4-Flash-0731} & 15.0 & {\color{dimtext}16.2} & 40.0 & {\color{dimtext}40.0} & 10.0 & {\color{dimtext}10.3} & 10.0 & {\color{dimtext}11.1} & 15.6 & {\color{dimtext}16.5} \\
\multirow{-10}{*}{\rotatebox[origin=c]{90}{\textbf{Claude Code}}} & \mdl{Claude Haiku 4.5} & 20.0 & {\color{dimtext}20.0} & 16.7 & {\color{dimtext}16.7} & 10.0 & {\color{dimtext}13.3} & 100.0 & {\color{dimtext}100.0} & 34.4 & {\color{dimtext}36.5} \\
\midrule
 & \mdl{Qwen-3.5-27B-AWQ-4bit} & 12.9 & {\color{dimtext}12.9} & 5.0 & {\color{dimtext}6.7} & 0.0 & {\color{dimtext}0.0} & 38.0 & {\color{dimtext}38.0} & 18.1 & {\color{dimtext}18.8} \\
 & \mdl{Qwen-3.5-4B} & 5.7 & {\color{dimtext}6.5} & 3.3 & {\color{dimtext}3.8} & 25.0 & {\color{dimtext}38.5} & 34.0 & {\color{dimtext}35.4} & 16.8 & {\color{dimtext}19.8} \\
 & \mdl{Qwen-3.5-9B} & 16.7 & {\color{dimtext}16.9} & 4.9 & {\color{dimtext}5.4} & 13.3 & {\color{dimtext}19.0} & 10.0 & {\color{dimtext}10.3} & 11.8 & {\color{dimtext}12.8} \\
 & \mdl{Qwen-3.5-122B-A10B-AWQ-4bit} & 56.7 & {\color{dimtext}58.6} & 20.0 & {\color{dimtext}20.0} & 5.0 & {\color{dimtext}5.0} & 0.0 & {\color{dimtext}0.0} & \textbf{24.4} & {\color{dimtext}24.7} \\
 & \mdl{Gemma-4-26B-A4B} & 0.0 & {\color{dimtext}0.0} & 7.5 & {\color{dimtext}15.8} & 23.3 & {\color{dimtext}33.3} & {\color{dimtext}---} & {\color{dimtext}---} & 12.5 & {\color{dimtext}23.8} \\
 & \mdl{Gemma-4-31B} & 15.0 & {\color{dimtext}16.2} & 10.0 & {\color{dimtext}13.3} & 16.7 & {\color{dimtext}18.5} & 20.0 & {\color{dimtext}21.1} & 15.5 & {\color{dimtext}17.3} \\
 & \mdl{Nemotron-3-Nano-30B-A3B} & 0.0 & {\color{dimtext}---} & 25.0 & {\color{dimtext}83.3} & 10.0 & {\color{dimtext}40.0} & 22.5 & {\color{dimtext}81.8} & 16.0 & {\color{dimtext}72.7} \\
 & \mdl{GPT-5.4-mini} & 10.0 & {\color{dimtext}10.5} & 10.0 & {\color{dimtext}10.0} & 18.0 & {\color{dimtext}22.5} & 5.0 & {\color{dimtext}5.6} & 13.0 & {\color{dimtext}14.9} \\
 & \mdl{DeepSeek-V4-Flash-0731} & 5.0 & {\color{dimtext}5.0} & {\color{dimtext}---} & {\color{dimtext}---} & 0.0 & {\color{dimtext}0.0} & 60.0 & {\color{dimtext}60.0} & 11.2 & {\color{dimtext}11.2} \\
\multirow{-10}{*}{\rotatebox[origin=c]{90}{\textbf{Codex}}} & \mdl{Claude Haiku 4.5} & 15.0 & {\color{dimtext}15.0} & {\color{dimtext}---} & {\color{dimtext}---} & 0.0 & {\color{dimtext}0.0} & 12.5 & {\color{dimtext}12.5} & 10.0 & {\color{dimtext}10.1} \\
\bottomrule
\end{tabular}
\end{adjustbox}
\end{table*}
\subsection{RQ3: Impact of Prompt Language}
\label{sec:rq3}

Developers worldwide instruct agents in their own languages. We therefore compare English and Traditional Chinese coding prompts across six agents using \mdl{Qwen-3.5-27B-AWQ-4bit}. \Cref{tab:rq2} reports the English results and \cref{tab:rq3} the Traditional Chinese results. The effect of prompt language varies across agents. Traditional Chinese prompts raise the overall gross ASR of OpenCode from 30.8\% to 36.9\% and of Claude Code from 25.0\% to 29.0\%, while mini-swe remains at 26.7\%. In contrast, Aider decreases substantially from 56.0\% to 37.0\%, followed by OpenHands (22.5\% to 20.0\%) and Codex (23.8\% to 22.5\%). For CWE-78, OpenCode increases from 32.0\% to 46.0\%, while Aider decreases from 100.0\% to 50.0\%. The attack thus remains effective across prompt languages, while the direction of the change depends on the agent framework. In the ten-LLM setting (Appendix~\ref{appdix:rq4}), the trend is clearer: Traditional Chinese prompts raise the overall gross ASR in 23 of 30 agent--model configurations, and the mean overall gross ASR from 22.3\% to 27.7\%.

\subsection{RQ4: Impact of Underlying Language Models}
\label{sec:rq4}

We next extend the evaluation to ten LLMs across OpenCode, Claude Code, and Codex with English prompts. As shown in \cref{tab:rq4}, ASRs vary substantially across models, even within the same agent framework. For OpenCode, overall gross ASRs range from 5.0\% (\mdl{GPT-5.4-mini}) to 75.0\% (\mdl{Gemma-4-31B}); for Claude Code, from 12.0\% (\mdl{GPT-5.4-mini}) to 33.3\% (\mdl{Claude Haiku 4.5}); and for Codex, from 8.8\% (\mdl{Claude Haiku 4.5}) to 25.0\% (\mdl{DeepSeek-V4-Flash-0731}). The same model can behave very differently across frameworks: \mdl{Gemma-4-31B} reaches 75.0\% with OpenCode but 30.0\% with Claude Code and 11.8\% with Codex. Susceptibility therefore depends on the interaction between the LLM and the agent framework, and the availability of qualifying bait also varies across models and CWE categories. The gap between gross and parseable ASR is also informative: with Codex and \mdl{Nemotron-3-Nano-30B-A3B}, the overall gross ASR is 11.0\% but the parseable ASR is 55.0\%, so although most completions cannot be analyzed, more than half of the analyzable ones contain the targeted weakness. Across CWEs, averaged over the configurations with qualifying bait, the mean gross ASR is highest for CWE-502 (28.2\%) and CWE-327 (26.8\%), followed by CWE-78 (18.8\%) and CWE-79 (11.9\%). CWE-79 also has the lowest bait yield, with no qualifying bait in 7 of the 30 configurations, compared with 1 for CWE-78, 3 for CWE-327, and 4 for CWE-502. Averaged across models, OpenCode (29.8\%) is more susceptible than Claude Code (21.9\%) and Codex (15.3\%). Results for the same setting with Traditional Chinese prompts are reported in Appendix~\ref{appdix:rq4}.

\subsection{RQ5: Cross-Model Transferability}
\label{sec:rq5}

In practice, the attacker does not know which LLM powers the victim agent. We therefore examine whether bait constructed with one model induces vulnerabilities when completed by another. \Cref{tab:cross_models_cwe78} reports cross-model transferability for CWE-78 across three agents and ten LLMs. Bait often transfers, and some bait is even more effective on other models than on its source model. For example, bait generated by \mdl{Qwen-3.5-4B} with OpenCode achieves a gross ASR of 20.0\% against itself but 100.0\% against \mdl{Qwen-3.5-122B-A10B-AWQ-4bit}. Similarly, bait generated by \mdl{GPT-5.4-mini} with Codex achieves 10.0\% against itself but 85.0\% against \mdl{DeepSeek-V4-Flash-0731}. Transferability varies across model pairs, and some combinations yield no successful attack. Some bait transfers broadly: bait generated by \mdl{Claude Haiku 4.5} with Codex induces CWE-78 on all ten completion models, with gross ASRs from 5.0\% to 55.0\%. Aggregated over all cross-model pairs with qualifying bait, the bait induces CWE-78 in at least one completion for 42 of 81 pairs (51.9\%) with OpenCode, 66 of 90 (73.3\%) with Claude Code, and 63 of 90 (70.0\%) with Codex; in 68 of these 261 pairs (26.1\%), the transferred bait is more effective than on its source model. Overall, identifier-based poisoning transfers across LLMs, so an attacker can construct bait with an accessible model and still compromise agents built on others. Appendix~\ref{appdix:more_cross_models} reports the corresponding results for CWE-79, CWE-327, and CWE-502.

\subsection{RQ6: Effectiveness of Prompt-Based Defense}
\label{sec:rq6}

A natural first-line defense is to remind the agent to write secure code. We append a single sentence to the English coding prompt instructing the agent to avoid generating code containing CWE vulnerabilities, and follow the RQ4 settings with ten LLMs and three agents. As shown in \cref{tab:rq6-1}, the effect of this defense varies considerably across agent--model configurations. Compared with the undefended English prompts, the overall gross ASR decreases in 13 configurations, increases in 15, and remains unchanged in two. For instance, OpenCode with \mdl{Gemma-4-31B} drops from 75.0\% to 30.0\%, whereas OpenCode with \mdl{Gemma-4-26B-A4B} rises from 47.5\% to 62.5\%, and Claude Code with \mdl{Qwen-3.5-27B-AWQ-4bit} rises from 25.0\% to 29.0\%. The reminder also leaves the most vulnerable configurations intact: Claude Code with \mdl{Claude Haiku 4.5} still yields 100.0\% on CWE-502, exactly as without the defense. Averaged over all 30 configurations, the reminder lowers the mean overall gross ASR only from 22.3\% to 19.2\%, and the mean for CWE-502 even rises from 28.2\% to 29.8\%. Generic security instructions therefore do not reliably prevent identifier-based poisoning: the poisoned names do not contradict the reminder, because they present the insecure choice as an established, trusted convention of the codebase.

\begin{table}[t]
\centering
\caption{Detection rates (\%) of targeted CWEs in poisoned (bait) and original (control) code generated using OpenCode with ten LLMs ($n=67$ pairs).}
\label{tab:model_tool_comparison}
\small
\setlength{\aboverulesep}{0pt}\setlength{\belowrulesep}{0pt}\setlength{\tabcolsep}{3.5pt}\renewcommand{\arraystretch}{1.1}
\begin{adjustbox}{max width=\linewidth}
\begin{tabular}{lcccc>{\columncolor{allcol}}c}
\toprule
\rowcolor{hdrgray}\textbf{Model} & \textbf{CWE-78} & \textbf{CWE-79} & \textbf{CWE-327} & \textbf{CWE-502} & \textbf{All} \\
\midrule
\mdl{Qwen-3.5-27B-AWQ-4bit} & 0.0 / 0.0 & 0.0 / 0.0 & 0.0 / 0.0 & 0.0 / 0.0 & \textbf{0.0 / 0.0} \\
\mdl{Qwen-3.5-4B} & 0.0 / 0.0 & 0.0 / 0.0 & {\color{dimtext}---} & {\color{dimtext}---} & \textbf{0.0 / 0.0} \\
\mdl{Qwen-3.5-9B} & 0.0 / 0.0 & 0.0 / 0.0 & 0.0 / 0.0 & 0.0 / 0.0 & \textbf{0.0 / 0.0} \\
\mdl{Qwen-3.5-122B-A10B-AWQ-4bit} & 0.0 / 0.0 & 0.0 / 0.0 & {\color{dimtext}---} & 0.0 / 0.0 & \textbf{0.0 / 0.0} \\
\mdl{Gemma-4-26B-A4B} & 0.0 / 0.0 & {\color{dimtext}---} & 0.0 / 0.0 & 0.0 / 0.0 & \textbf{0.0 / 0.0} \\
\mdl{Gemma-4-31B} & 0.0 / 0.0 & {\color{dimtext}---} & 0.0 / 0.0 & 0.0 / 0.0 & \textbf{0.0 / 0.0} \\
\mdl{Nemotron-3-Nano-30B-A3B} & 0.0 / 0.0 & 0.0 / 0.0 & 0.0 / 0.0 & 0.0 / 0.0 & \textbf{0.0 / 0.0} \\
\mdl{GPT-5.4-mini} & 0.0 / 0.0 & {\color{dimtext}---} & 0.0 / 0.0 & 0.0 / 0.0 & \textbf{0.0 / 0.0} \\
\mdl{DeepSeek-V4-Flash-0731} & 0.0 / 0.0 & {\color{dimtext}---} & 0.0 / 0.0 & 0.0 / 0.0 & \textbf{0.0 / 0.0} \\
\mdl{Claude Haiku 4.5} & {\color{dimtext}---} & {\color{dimtext}---} & 0.0 / 0.0 & {\color{dimtext}---} & \textbf{0.0 / 0.0} \\
\midrule
\rowcolor{hdrgray}\textbf{Overall} & \textbf{0.0 / 0.0} & \textbf{0.0 / 0.0} & \textbf{0.0 / 0.0} & \textbf{0.0 / 0.0} & \textbf{0.0 / 0.0} \\
\bottomrule
\end{tabular}
\end{adjustbox}
\par\smallskip{\footnotesize\raggedright Each cell reports bait/control detection rates and holds for all seven tools (\mdl{CodeGemma}, \mdl{CodeAstra-7B}, \mdl{GPT-5.4-mini}, Bandit, Semgrep, CodeShield, and CodeQL), except CodeShield on CWE-79, which has no applicable rule in any row. ---: no qualifying bait code.\par}
\end{table}
\subsection{RQ7: Detectability of Poisoned Source Code}
\label{sec:rq7}

We then ask whether existing security analyzers can flag poisoned third-party files before a victim agent processes them. We evaluate three AI-based analyzers, namely \mdl{CodeGemma} (a code-specialized LLM), \mdl{CodeAstra-7B} (an LLM fine-tuned for vulnerability detection), and \mdl{GPT-5.4-mini} prompted as a security reviewer, as well as four static analyzers: Bandit (an AST-based Python security linter), Semgrep (rule-based pattern matching), CodeShield (Meta's detector for insecure LLM-generated code), and CodeQL (query-based semantic and data-flow analysis). For each case, we compare the results on the poisoned file with those on its clean counterpart, focusing on whether the targeted CWE is reported.

As shown in \cref{tab:model_tool_comparison}, all seven tools achieve a 0.0\% detection rate for the targeted CWEs in both bait and control code produced with OpenCode across all ten LLMs and all CWE categories for which qualifying bait and an applicable rule exist. Because \nametrap{} modifies only identifiers and introduces no vulnerable statement, present-code vulnerability detectors have nothing to report, even though the file reliably induces vulnerable code once an agent completes it. The same holds for bait constructed with Claude Code and Codex (Appendix~\ref{appdix:detection_results}). This blind spot is structural rather than tool-specific: rule-based analyzers match vulnerable statements or data flows, which the bait does not contain, while LLM-based reviewers, although they read identifiers, assess whether the present code is vulnerable rather than whether its names would mislead a future implementer.

\begin{table}[t]
\centering
\caption{Comparison of attack success rates (parseable ASR, \%) between XOXO and \nametrap{} across four CWE categories. \nametrap{} is evaluated with six coding agents, while XOXO does not distinguish between coding agents.}
\label{tab:xoxo_comparison}
\small
\setlength{\aboverulesep}{0pt}\setlength{\belowrulesep}{0pt}\setlength{\tabcolsep}{4pt}\renewcommand{\arraystretch}{1.12}
\begin{adjustbox}{max width=\columnwidth}
\begin{tabular}{llcccc>{\columncolor{allcol}}c}
\toprule
\rowcolor{hdrgray}\textbf{Method} & \textbf{Coding agent} & \textbf{CWE-78} & \textbf{CWE-79} & \textbf{CWE-327} & \textbf{CWE-502} & \textbf{Overall} \\
\midrule
XOXO & {\color{dimtext}N/A} & 0.0 & 0.0 & 10.0 & 3.3 & 2.1 \\
\midrule
\rowcolor{oursrow} & OpenCode & 32.0 & 30.0 & 60.0 & 24.0 & 30.8 \\
\rowcolor{oursrow} & Claude Code & 11.7 & 0.0 & 70.0 & 55.0 & 25.0 \\
\rowcolor{oursrow} & Codex & 15.7 & 37.5 & 0.0 & 42.0 & 24.7 \\
\rowcolor{oursrow} & mini-swe & 30.0 & 10.0 & 40.0 & 55.0 & 26.7 \\
\rowcolor{oursrow} & Aider & 100.0 & 63.3 & 60.0 & 27.5 & 56.0 \\
\rowcolor{oursrow}\multirow{-6}{*}{\textbf{\nametrap}} & OpenHands & 33.3 & 5.0 & 29.4 & 20.0 & 23.4 \\
\bottomrule
\end{tabular}
\end{adjustbox}
\end{table}
\subsection{RQ8: Comparison with XOXO}
\label{sec:rq8}

\Cref{tab:xoxo_comparison} compares \nametrap{} with XOXO~\cite{storek2026xoxo}, the closest prior identifier-level context-poisoning attack, on the same four CWE categories. XOXO is evaluated with 200 queries and does not distinguish between coding agents, so it is reported as a single row. XOXO achieves an overall ASR of only 2.1\% and no success on CWE-78 or CWE-79, whereas \nametrap{} achieves 23.4\% to 56.0\% overall across all six agents. Even on CWE-327, XOXO's strongest category (10.0\%), \nametrap{} reaches 70.0\% with Claude Code and 60.0\% with OpenCode and Aider. The gap reflects \nametrap's design: identifier choices are conditioned on the trust assumption underlying each target CWE and verified through probe completions before release.

\section{Discussion}
\label{sec:discussion}

\textbf{Why identifier poisoning works.} In an incomplete file, the surrounding names are the main evidence an agent has about the intended behavior of the missing code. Names that denote trusted commands, prevalidated output, legacy-compatible digests, or trusted serialized state encode a trust assumption, and an agent that follows the conventions of the codebase, which is normally desirable behavior, consequently omits sanitization, escaping, or safe APIs. Because the bait is valid and functionally identical code, it contains nothing that instruction-oriented filters or present-code vulnerability scanners are designed to flag (RQ7).

\textbf{Why some CWEs are easier to induce.} The per-CWE results (RQ4) follow how much the poisoned names must justify. For CWE-502 and CWE-327, the vulnerable choice is a single, idiomatic standard-library call, such as pickle-based deserialization or an MD5 or SHA-1 digest, which becomes the most direct implementation once the surrounding names assert that the data is trusted or that legacy compatibility is required. For CWE-78, names that mark a command as trusted make shell-based execution appear consistent with the codebase. For CWE-79, the agent must actively omit escaping when constructing output, which overrides a strong default in Web code; this explains both the lower ASR and the lower bait yield for this category.

\textbf{Implications for the Web ecosystem.} Public code-hosting platforms and package registries already serve as distribution channels for malicious software~\cite{zimmermann2019npm}. \nametrap{} shows that, once agents consume this content, an attacker no longer needs to ship malicious code; shipping misleading names suffices. Because bait transfers across LLMs (RQ5) and remains effective under different prompt languages (RQ3), a single published file can affect agents built on models the attacker never queried. Identifier-only bait also suits the channels through which agents acquire code on the Web. Because the bait is functionally identical to the original, it can be submitted as a cosmetic naming-cleanup pull request, distributed as a fork or mirror of a popular project, or published as example code in a tutorial, gist, or Q\&A answer that agents retrieve through Web search. Unlike a malicious dependency, such bait passes functional tests and leaves no behavioral trace until an agent extends it.

\textbf{Mitigations.} RQ6 and RQ7 show that defenses must shift from inspecting the ingested artifact to inspecting the generated code and its provenance. Agent frameworks should (i) treat identifiers in third-party code as untrusted input rather than as specifications, (ii) run security analysis on every agent-generated change before it is committed, since the targeted weaknesses become detectable once written, and (iii) record which external files influenced each generated edit so that reviewers can audit changes that depend on third-party code.

\textbf{Ethical considerations.} All bait files were constructed and evaluated in isolated local environments. No poisoned file was published to a public repository or package registry, and the study involved no human subjects. All source projects, coding agents, and language models used in this study are publicly available.

\afterpage{\setcounter{dbltopnumber}{2}%
\begin{table*}[t]
\centering
\caption{Attack success rates (ASR, \%) across ten LLMs and three coding agents using Traditional Chinese prompts. Bold: highest overall gross ASR per agent.}
\label{tab:rq3-1}
\footnotesize
\setlength{\aboverulesep}{0pt}\setlength{\belowrulesep}{0pt}\setlength{\tabcolsep}{4.5pt}\renewcommand{\arraystretch}{0.9}
\begin{adjustbox}{max width=\textwidth}
\begin{tabular}{cl*{8}{c}>{\columncolor{allcol}}c>{\columncolor{allcol}}c}
\toprule
\rowcolor{hdrgray} & & \multicolumn{2}{c}{\textbf{CWE-78}} & \multicolumn{2}{c}{\textbf{CWE-79}} & \multicolumn{2}{c}{\textbf{CWE-327}} & \multicolumn{2}{c}{\textbf{CWE-502}} & \multicolumn{2}{c}{\textbf{All}} \\
\rowcolor{hdrgray}\multirow{-2}{*}{\textbf{Agent}} & \multirow{-2}{*}{\textbf{Model}} & G & P & G & P & G & P & G & P & G & P \\
\midrule
 & \mdl{Qwen-3.5-27B-AWQ-4bit} & 46.0 & {\color{dimtext}46.0} & 20.0 & {\color{dimtext}20.0} & 70.0 & {\color{dimtext}70.0} & 28.0 & {\color{dimtext}28.0} & 36.9 & {\color{dimtext}36.9} \\
 & \mdl{Qwen-3.5-4B} & 50.0 & {\color{dimtext}50.0} & 0.0 & {\color{dimtext}0.0} & {\color{dimtext}---} & {\color{dimtext}---} & {\color{dimtext}---} & {\color{dimtext}---} & 25.0 & {\color{dimtext}25.0} \\
 & \mdl{Qwen-3.5-9B} & 20.0 & {\color{dimtext}20.0} & 10.0 & {\color{dimtext}10.0} & 70.0 & {\color{dimtext}70.0} & 85.0 & {\color{dimtext}85.0} & 54.0 & {\color{dimtext}54.0} \\
 & \mdl{Qwen-3.5-122B-A10B-AWQ-4bit} & 36.7 & {\color{dimtext}36.7} & 38.0 & {\color{dimtext}38.0} & {\color{dimtext}---} & {\color{dimtext}---} & 53.3 & {\color{dimtext}53.3} & 41.8 & {\color{dimtext}41.8} \\
 & \mdl{Gemma-4-26B-A4B} & 0.0 & {\color{dimtext}0.0} & {\color{dimtext}---} & {\color{dimtext}---} & 80.0 & {\color{dimtext}88.9} & 60.0 & {\color{dimtext}60.0} & 50.0 & {\color{dimtext}51.3} \\
 & \mdl{Gemma-4-31B} & 60.0 & {\color{dimtext}60.0} & {\color{dimtext}---} & {\color{dimtext}---} & 100.0 & {\color{dimtext}100.0} & 65.0 & {\color{dimtext}65.0} & \textbf{72.5} & {\color{dimtext}72.5} \\
 & \mdl{Nemotron-3-Nano-30B-A3B} & 6.7 & {\color{dimtext}7.1} & 0.0 & {\color{dimtext}0.0} & 13.3 & {\color{dimtext}14.8} & 10.0 & {\color{dimtext}10.0} & 8.2 & {\color{dimtext}8.6} \\
 & \mdl{GPT-5.4-mini} & 0.0 & {\color{dimtext}0.0} & {\color{dimtext}---} & {\color{dimtext}---} & 10.0 & {\color{dimtext}10.0} & 20.0 & {\color{dimtext}20.0} & 10.0 & {\color{dimtext}10.0} \\
 & \mdl{DeepSeek-V4-Flash-0731} & 5.0 & {\color{dimtext}5.0} & {\color{dimtext}---} & {\color{dimtext}---} & 30.0 & {\color{dimtext}30.0} & 30.0 & {\color{dimtext}30.0} & 10.0 & {\color{dimtext}10.0} \\
\multirow{-10}{*}{\rotatebox[origin=c]{90}{\textbf{OpenCode}}} & \mdl{Claude Haiku 4.5} & {\color{dimtext}---} & {\color{dimtext}---} & {\color{dimtext}---} & {\color{dimtext}---} & 50.0 & {\color{dimtext}50.0} & {\color{dimtext}---} & {\color{dimtext}---} & 50.0 & {\color{dimtext}50.0} \\
\midrule
 & \mdl{Qwen-3.5-27B-AWQ-4bit} & 20.0 & {\color{dimtext}20.0} & 0.0 & {\color{dimtext}0.0} & 70.0 & {\color{dimtext}70.0} & 50.0 & {\color{dimtext}50.0} & 29.0 & {\color{dimtext}29.0} \\
 & \mdl{Qwen-3.5-4B} & 7.5 & {\color{dimtext}8.6} & 13.3 & {\color{dimtext}16.0} & 60.0 & {\color{dimtext}60.0} & 36.7 & {\color{dimtext}41.5} & 25.0 & {\color{dimtext}28.5} \\
 & \mdl{Qwen-3.5-9B} & 27.1 & {\color{dimtext}27.9} & 18.0 & {\color{dimtext}18.4} & 90.0 & {\color{dimtext}90.0} & 25.0 & {\color{dimtext}31.2} & 27.6 & {\color{dimtext}29.6} \\
 & \mdl{Qwen-3.5-122B-A10B-AWQ-4bit} & 45.0 & {\color{dimtext}45.0} & 55.0 & {\color{dimtext}55.0} & {\color{dimtext}---} & {\color{dimtext}---} & {\color{dimtext}---} & {\color{dimtext}---} & \textbf{50.0} & {\color{dimtext}50.0} \\
 & \mdl{Gemma-4-26B-A4B} & 10.0 & {\color{dimtext}10.0} & 6.7 & {\color{dimtext}9.5} & 27.5 & {\color{dimtext}30.6} & 5.0 & {\color{dimtext}5.0} & 13.8 & {\color{dimtext}15.4} \\
 & \mdl{Gemma-4-31B} & 30.0 & {\color{dimtext}30.0} & 56.5 & {\color{dimtext}57.8} & 26.7 & {\color{dimtext}26.7} & 10.0 & {\color{dimtext}10.0} & 36.2 & {\color{dimtext}36.5} \\
 & \mdl{Nemotron-3-Nano-30B-A3B} & 15.0 & {\color{dimtext}18.8} & 0.0 & {\color{dimtext}0.0} & 13.3 & {\color{dimtext}16.0} & 65.0 & {\color{dimtext}86.7} & 23.0 & {\color{dimtext}30.3} \\
 & \mdl{GPT-5.4-mini} & 5.0 & {\color{dimtext}5.3} & 26.7 & {\color{dimtext}28.6} & 15.0 & {\color{dimtext}15.0} & 30.0 & {\color{dimtext}34.6} & 21.0 & {\color{dimtext}22.6} \\
 & \mdl{DeepSeek-V4-Flash-0731} & 25.0 & {\color{dimtext}26.3} & 20.0 & {\color{dimtext}20.0} & 20.0 & {\color{dimtext}22.2} & 10.0 & {\color{dimtext}10.0} & 21.1 & {\color{dimtext}22.4} \\
\multirow{-10}{*}{\rotatebox[origin=c]{90}{\textbf{Claude Code}}} & \mdl{Claude Haiku 4.5} & 40.0 & {\color{dimtext}40.0} & 16.7 & {\color{dimtext}16.7} & 0.0 & {\color{dimtext}0.0} & 100.0 & {\color{dimtext}100.0} & 36.7 & {\color{dimtext}40.2} \\
\midrule
 & \mdl{Qwen-3.5-27B-AWQ-4bit} & 14.3 & {\color{dimtext}14.3} & 25.0 & {\color{dimtext}29.4} & 5.0 & {\color{dimtext}5.3} & 40.0 & {\color{dimtext}40.0} & 22.5 & {\color{dimtext}23.1} \\
 & \mdl{Qwen-3.5-4B} & 12.9 & {\color{dimtext}13.6} & 13.3 & {\color{dimtext}14.8} & 7.5 & {\color{dimtext}10.7} & 36.0 & {\color{dimtext}40.9} & 17.9 & {\color{dimtext}20.6} \\
 & \mdl{Qwen-3.5-9B} & 21.1 & {\color{dimtext}21.6} & 7.9 & {\color{dimtext}8.3} & 3.3 & {\color{dimtext}3.6} & 20.0 & {\color{dimtext}21.4} & 14.6 & {\color{dimtext}15.2} \\
 & \mdl{Qwen-3.5-122B-A10B-AWQ-4bit} & 46.7 & {\color{dimtext}48.3} & 0.0 & {\color{dimtext}0.0} & 5.0 & {\color{dimtext}5.0} & 5.0 & {\color{dimtext}5.0} & 17.8 & {\color{dimtext}18.0} \\
 & \mdl{Gemma-4-26B-A4B} & 0.0 & {\color{dimtext}0.0} & 2.5 & {\color{dimtext}5.9} & 63.3 & {\color{dimtext}70.4} & {\color{dimtext}---} & {\color{dimtext}---} & 25.0 & {\color{dimtext}42.6} \\
 & \mdl{Gemma-4-31B} & 25.0 & {\color{dimtext}28.6} & 15.0 & {\color{dimtext}20.0} & 20.0 & {\color{dimtext}28.6} & 5.0 & {\color{dimtext}5.9} & 18.2 & {\color{dimtext}22.7} \\
 & \mdl{Nemotron-3-Nano-30B-A3B} & 0.0 & {\color{dimtext}---} & 5.0 & {\color{dimtext}50.0} & 5.0 & {\color{dimtext}33.3} & 15.0 & {\color{dimtext}75.0} & 8.0 & {\color{dimtext}61.5} \\
 & \mdl{GPT-5.4-mini} & 25.0 & {\color{dimtext}29.4} & 10.0 & {\color{dimtext}11.1} & 20.0 & {\color{dimtext}22.7} & 5.0 & {\color{dimtext}6.2} & 17.0 & {\color{dimtext}19.8} \\
 & \mdl{DeepSeek-V4-Flash-0731} & 28.3 & {\color{dimtext}28.3} & {\color{dimtext}---} & {\color{dimtext}---} & 20.0 & {\color{dimtext}22.2} & 80.0 & {\color{dimtext}80.0} & \textbf{33.8} & {\color{dimtext}34.2} \\
\multirow{-10}{*}{\rotatebox[origin=c]{90}{\textbf{Codex}}} & \mdl{Claude Haiku 4.5} & 5.0 & {\color{dimtext}5.0} & {\color{dimtext}---} & {\color{dimtext}---} & 15.0 & {\color{dimtext}20.0} & 20.0 & {\color{dimtext}20.0} & 15.0 & {\color{dimtext}16.0} \\
\bottomrule
\end{tabular}
\end{adjustbox}
\end{table*}
\begin{table*}[t]
\begin{minipage}[t]{0.485\textwidth}
\centering
\caption{Detection results for bait code generated with Claude Code across AI-based and traditional static analysis tools.}
\label{tab:detection_claude}
\small
\setlength{\aboverulesep}{0pt}\setlength{\belowrulesep}{0pt}\setlength{\tabcolsep}{3.5pt}\renewcommand{\arraystretch}{1.1}
\begin{adjustbox}{max width=\linewidth}
\begin{tabular}{lcccc>{\columncolor{allcol}}c}
\toprule
\rowcolor{hdrgray}\textbf{Model} & \textbf{CWE-78} & \textbf{CWE-79} & \textbf{CWE-327} & \textbf{CWE-502} & \textbf{All} \\
\midrule
\mdl{Qwen-3.5-27B-AWQ-4bit} & 0.0 / 0.0$^{a}$ & 0.0 / 0.0 & 0.0 / 0.0 & 0.0 / 0.0 & \textbf{0.0 / 0.0$^{b}$} \\
\mdl{Qwen-3.5-4B} & 0.0 / 0.0 & 0.0 / 0.0 & 0.0 / 0.0 & 0.0 / 0.0 & \textbf{0.0 / 0.0} \\
\mdl{Qwen-3.5-9B} & 0.0 / 0.0 & 0.0 / 0.0 & 0.0 / 0.0 & 0.0 / 0.0 & \textbf{0.0 / 0.0} \\
\mdl{Qwen-3.5-122B-A10B-AWQ-4bit} & 0.0 / 0.0 & 0.0 / 0.0 & {\color{dimtext}---} & {\color{dimtext}---} & \textbf{0.0 / 0.0} \\
\mdl{Gemma-4-26B-A4B} & 0.0 / 0.0 & 0.0 / 0.0 & 0.0 / 0.0 & 0.0 / 0.0 & \textbf{0.0 / 0.0} \\
\mdl{Gemma-4-31B} & 0.0 / 0.0 & 0.0 / 0.0 & 0.0 / 0.0 & 0.0 / 0.0 & \textbf{0.0 / 0.0} \\
\mdl{Nemotron-3-Nano-30B-A3B} & 0.0 / 0.0 & 0.0 / 0.0 & 0.0 / 0.0 & 0.0 / 0.0 & \textbf{0.0 / 0.0} \\
\mdl{GPT-5.4-mini} & 0.0 / 0.0 & 0.0 / 0.0 & 0.0 / 0.0 & 0.0 / 0.0 & \textbf{0.0 / 0.0} \\
\mdl{DeepSeek-V4-Flash-0731} & 0.0 / 0.0 & 0.0 / 0.0 & 0.0 / 0.0 & 0.0 / 0.0 & \textbf{0.0 / 0.0} \\
\mdl{Claude Haiku 4.5} & 0.0 / 0.0 & 0.0 / 0.0 & 0.0 / 0.0 & 0.0 / 0.0 & \textbf{0.0 / 0.0} \\
\midrule
\rowcolor{hdrgray}\textbf{Overall} & \textbf{0.0 / 0.0$^{c}$} & \textbf{0.0 / 0.0} & \textbf{0.0 / 0.0} & \textbf{0.0 / 0.0} & \textbf{0.0 / 0.0$^{d}$} \\
\bottomrule
\end{tabular}
\end{adjustbox}
\par\smallskip{\footnotesize\raggedright Each cell reports bait/control detection rates and holds for all seven tools (\mdl{CodeGemma}, \mdl{CodeAstra-7B}, \mdl{GPT-5.4-mini}, Bandit, Semgrep, CodeShield, and CodeQL), except CodeShield on CWE-79, which has no applicable rule in any row. ---: no qualifying bait code. Exceptions (\mdl{CodeGemma}, bait/control): $^{a}$0.0 / 16.7; $^{b}$0.0 / 10.0; $^{c}$0.0 / 2.7; $^{d}$0.0 / 0.9.\par}
\end{minipage}\hfill
\begin{minipage}[t]{0.485\textwidth}
\centering
\caption{Detection results for bait code generated with Codex CLI across AI-based and traditional static analysis tools.}
\label{tab:detection_codex}
\small
\setlength{\aboverulesep}{0pt}\setlength{\belowrulesep}{0pt}\setlength{\tabcolsep}{3.5pt}\renewcommand{\arraystretch}{1.1}
\begin{adjustbox}{max width=\linewidth}
\begin{tabular}{lcccc>{\columncolor{allcol}}c}
\toprule
\rowcolor{hdrgray}\textbf{Model} & \textbf{CWE-78} & \textbf{CWE-79} & \textbf{CWE-327} & \textbf{CWE-502} & \textbf{All} \\
\midrule
\mdl{Qwen-3.5-27B-AWQ-4bit} & 0.0 / 0.0 & 0.0 / 0.0 & 0.0 / 0.0 & 0.0 / 0.0 & \textbf{0.0 / 0.0} \\
\mdl{Qwen-3.5-4B} & 0.0 / 0.0 & 0.0 / 0.0 & 0.0 / 0.0 & 0.0 / 0.0 & \textbf{0.0 / 0.0} \\
\mdl{Qwen-3.5-9B} & 0.0 / 0.0 & 0.0 / 0.0 & 0.0 / 0.0 & 0.0 / 0.0 & \textbf{0.0 / 0.0} \\
\mdl{Qwen-3.5-122B-A10B-AWQ-4bit} & 0.0 / 0.0 & 0.0 / 0.0 & 0.0 / 0.0 & 0.0 / 0.0 & \textbf{0.0 / 0.0} \\
\mdl{Gemma-4-26B-A4B} & 0.0 / 0.0 & 0.0 / 0.0 & 0.0 / 0.0 & {\color{dimtext}---} & \textbf{0.0 / 0.0} \\
\mdl{Gemma-4-31B} & 0.0 / 0.0 & 0.0 / 0.0 & 0.0 / 0.0 & 0.0 / 0.0 & \textbf{0.0 / 0.0} \\
\mdl{Nemotron-3-Nano-30B-A3B} & 0.0 / 0.0 & 0.0 / 0.0 & 0.0 / 0.0 & 0.0 / 0.0 & \textbf{0.0 / 0.0} \\
\mdl{GPT-5.4-mini} & 0.0 / 0.0 & 0.0 / 0.0 & 0.0 / 0.0 & 0.0 / 0.0 & \textbf{0.0 / 0.0} \\
\mdl{DeepSeek-V4-Flash-0731} & 0.0 / 0.0 & {\color{dimtext}---} & 0.0 / 0.0 & 0.0 / 0.0 & \textbf{0.0 / 0.0} \\
\mdl{Claude Haiku 4.5} & 0.0 / 0.0 & {\color{dimtext}---} & 0.0 / 0.0 & 0.0 / 0.0 & \textbf{0.0 / 0.0} \\
\midrule
\rowcolor{hdrgray}\textbf{Overall} & \textbf{0.0 / 0.0} & \textbf{0.0 / 0.0} & \textbf{0.0 / 0.0} & \textbf{0.0 / 0.0} & \textbf{0.0 / 0.0} \\
\bottomrule
\end{tabular}
\end{adjustbox}
\par\smallskip{\footnotesize\raggedright Each cell reports bait/control detection rates and holds for all seven tools (\mdl{CodeGemma}, \mdl{CodeAstra-7B}, \mdl{GPT-5.4-mini}, Bandit, Semgrep, CodeShield, and CodeQL), except CodeShield on CWE-79, which has no applicable rule in any row. ---: no qualifying bait code.\par}
\end{minipage}
\end{table*}

}
\section{Conclusion}
\label{sec:conclusion}

We presented \nametrap, an inference-time source-code poisoning attack that steers AI coding agents toward attacker-chosen vulnerabilities solely by renaming identifiers in third-party files. Across six coding agents, ten LLMs, and four CWE categories, identifier-only bait induces targeted vulnerabilities from screened clean baselines, outperforms random renaming, neutral renaming, and XOXO, transfers across models, survives a prompt-level security reminder, and evades seven AI-based and static analyzers. As coding agents increasingly build software from code retrieved from the Web, the semantics of untrusted identifiers must be treated as part of their attack surface. Future work includes extending \nametrap{} to other programming languages and CWE classes, and developing identifier-aware defenses that reason explicitly about the trust assumptions encoded in names.

\bibliographystyle{ACM-Reference-Format}
\bibliography{samples}

\appendix

\begin{table*}[t]
\centering
\caption{Cross-model transferability of bait code on CWE-79 (cross-site scripting). Same format as \cref{tab:cross_models_cwe78}.}
\label{tab:matrix_all_agents_079}
\scriptsize
\setlength{\aboverulesep}{0pt}\setlength{\belowrulesep}{0pt}\setlength{\tabcolsep}{3pt}\renewcommand{\arraystretch}{1.08}
\begin{adjustbox}{max width=\textwidth}
\begin{tabular}{cl*{10}{c}}
\toprule
\rowcolor{hdrgray}\rule{0pt}{2.4ex} & & \multicolumn{10}{c}{\textbf{Task-completion model}} \\
\cmidrule(lr){3-12}
\rowcolor{hdrgray} & & \texttt{Qwen-3.5-} & \texttt{Qwen-3.5-} & \texttt{Qwen-3.5-} & \texttt{Qwen-3.5-122B-} & \texttt{Gemma-4-} & \texttt{Gemma-4-} & \texttt{Nemotron-3-} & \texttt{GPT-5.4-} & \texttt{DeepSeek-V4-} & \texttt{Claude} \\
\rowcolor{hdrgray}\multirow{-3}{*}{\textbf{Agent}} & \multirow{-3}{*}{\textbf{Bait model}} & \texttt{27B-AWQ-4bit} & \texttt{4B} & \texttt{9B} & \texttt{A10B-AWQ-4bit} & \texttt{26B-A4B} & \texttt{31B} & \texttt{Nano-30B-A3B} & \texttt{mini} & \texttt{Flash-0731} & \texttt{Haiku 4.5} \\
\midrule
 & \mdl{Qwen-3.5-27B-AWQ-4bit} & \cellcolor{heat!13}\textbf{30.0} / 30.0 & 0.0 / 0.0 & 0.0 / 0.0 & \cellcolor{heat!4}10.0 / 10.0 & 0.0 / 0.0 & 0.0 / 0.0 & 0.0 / 0.0 & \cellcolor{heat!15}35.0 / 35.0 & \cellcolor{heat!40}95.0 / 95.0 & \cellcolor{heat!27}65.0 / 65.0 \\
 & \mdl{Qwen-3.5-4B} & 0.0 / 0.0 & \cellcolor{heat!4}\textbf{10.0} / 10.0 & 0.0 / 0.0 & \cellcolor{heat!13}30.0 / 30.0 & 0.0 / 0.0 & 0.0 / 0.0 & 0.0 / 0.0 & \cellcolor{heat!8}20.0 / 20.0 & \cellcolor{heat!42}100.0 / 100.0 & \cellcolor{heat!17}40.0 / 40.0 \\
 & \mdl{Qwen-3.5-9B} & \cellcolor{heat!42}100.0 / 100.0 & 0.0 / 0.0 & \cellcolor{heat!4}\textbf{10.0} / 10.0 & \cellcolor{heat!29}70.0 / 70.0 & 0.0 / 0.0 & 0.0 / 0.0 & 0.0 / 0.0 & \cellcolor{heat!17}40.0 / 40.0 & \cellcolor{heat!38}90.0 / 90.0 & \cellcolor{heat!42}100.0 / 100.0 \\
 & \mdl{Qwen-3.5-122B-A10B-AWQ-4bit} & \cellcolor{heat!41}98.0 / 98.0 & 0.0 / 0.0 & \cellcolor{heat!4}6.0 / 6.0 & \cellcolor{heat!12}\textbf{28.0} / 28.0 & 0.0 / 0.0 & 0.0 / 0.0 & 0.0 / 0.0 & \cellcolor{heat!18}42.0 / 42.0 & \cellcolor{heat!40}96.0 / 96.0 & \cellcolor{heat!39}94.0 / 94.0 \\
 & \mdl{Gemma-4-26B-A4B} & {\color{dimtext}---} & {\color{dimtext}---} & {\color{dimtext}---} & {\color{dimtext}---} & {\color{dimtext}---} & {\color{dimtext}---} & {\color{dimtext}---} & {\color{dimtext}---} & {\color{dimtext}---} & {\color{dimtext}---} \\
 & \mdl{Gemma-4-31B} & {\color{dimtext}---} & {\color{dimtext}---} & {\color{dimtext}---} & {\color{dimtext}---} & {\color{dimtext}---} & {\color{dimtext}---} & {\color{dimtext}---} & {\color{dimtext}---} & {\color{dimtext}---} & {\color{dimtext}---} \\
 & \mdl{Nemotron-3-Nano-30B-A3B} & \cellcolor{heat!42}100.0 / 100.0 & 0.0 / 0.0 & \cellcolor{heat!13}30.0 / 30.0 & \cellcolor{heat!21}50.0 / 50.0 & \cellcolor{heat!21}50.0 / 50.0 & \cellcolor{heat!21}50.0 / 50.0 & \textbf{0.0} / 0.0 & \cellcolor{heat!25}60.0 / 60.0 & \cellcolor{heat!38}90.0 / 90.0 & \cellcolor{heat!42}100.0 / 100.0 \\
 & \mdl{GPT-5.4-mini} & {\color{dimtext}---} & {\color{dimtext}---} & {\color{dimtext}---} & {\color{dimtext}---} & {\color{dimtext}---} & {\color{dimtext}---} & {\color{dimtext}---} & {\color{dimtext}---} & {\color{dimtext}---} & {\color{dimtext}---} \\
 & \mdl{DeepSeek-V4-Flash-0731} & {\color{dimtext}---} & {\color{dimtext}---} & {\color{dimtext}---} & {\color{dimtext}---} & {\color{dimtext}---} & {\color{dimtext}---} & {\color{dimtext}---} & {\color{dimtext}---} & {\color{dimtext}---} & {\color{dimtext}---} \\
\multirow{-10}{*}{\rotatebox[origin=c]{90}{\textbf{OpenCode}}} & \mdl{Claude Haiku 4.5} & {\color{dimtext}---} & {\color{dimtext}---} & {\color{dimtext}---} & {\color{dimtext}---} & {\color{dimtext}---} & {\color{dimtext}---} & {\color{dimtext}---} & {\color{dimtext}---} & {\color{dimtext}---} & {\color{dimtext}---} \\
\midrule
 & \mdl{Qwen-3.5-27B-AWQ-4bit} & \textbf{0.0} / 0.0 & \cellcolor{heat!4}10.0 / 10.0 & 0.0 / 0.0 & \cellcolor{heat!13}30.0 / 30.0 & 0.0 / 0.0 & 0.0 / {\color{dimtext}---} & 0.0 / 0.0 & 0.0 / 0.0 & 0.0 / 0.0 & \cellcolor{heat!29}70.0 / 70.0 \\
 & \mdl{Qwen-3.5-4B} & \cellcolor{heat!8}20.0 / 20.0 & \cellcolor{heat!4}\textbf{6.7} / 7.1 & \cellcolor{heat!6}13.3 / 13.8 & \cellcolor{heat!10}23.3 / 23.3 & 0.0 / 0.0 & \cellcolor{heat!4}10.0 / 12.5 & 0.0 / 0.0 & \cellcolor{heat!27}63.3 / 67.9 & \cellcolor{heat!31}73.3 / 75.9 & \cellcolor{heat!20}46.7 / 46.7 \\
 & \mdl{Qwen-3.5-9B} & \cellcolor{heat!33}78.0 / 78.0 & \cellcolor{heat!4}6.0 / 6.1 & \cellcolor{heat!4}\textbf{6.0} / 6.2 & \cellcolor{heat!27}64.0 / 64.0 & \cellcolor{heat!4}6.0 / 6.5 & \cellcolor{heat!4}6.0 / 6.0 & \cellcolor{heat!4}10.0 / 11.6 & \cellcolor{heat!35}84.0 / 84.0 & \cellcolor{heat!29}68.0 / 68.0 & \cellcolor{heat!41}98.0 / 98.0 \\
 & \mdl{Qwen-3.5-122B-A10B-AWQ-4bit} & \cellcolor{heat!29}70.0 / 70.0 & \cellcolor{heat!4}10.0 / 16.7 & \cellcolor{heat!4}10.0 / 10.0 & \cellcolor{heat!8}\textbf{20.0} / 20.0 & \cellcolor{heat!4}10.0 / 10.0 & \cellcolor{heat!4}5.0 / 5.9 & 0.0 / 0.0 & \cellcolor{heat!38}90.0 / 90.0 & \cellcolor{heat!29}70.0 / 70.0 & \cellcolor{heat!29}70.0 / 70.0 \\
 & \mdl{Gemma-4-26B-A4B} & \cellcolor{heat!17}40.0 / 40.0 & \cellcolor{heat!4}3.3 / 3.7 & \cellcolor{heat!4}6.7 / 7.4 & \cellcolor{heat!24}56.7 / 56.7 & \cellcolor{heat!4}\textbf{3.3} / 4.8 & \cellcolor{heat!6}13.3 / 15.4 & 0.0 / 0.0 & \cellcolor{heat!22}53.3 / 57.1 & \cellcolor{heat!34}80.0 / 85.7 & \cellcolor{heat!29}70.0 / 70.0 \\
 & \mdl{Gemma-4-31B} & \cellcolor{heat!30}72.3 / 72.3 & \cellcolor{heat!4}4.8 / 5.0 & \cellcolor{heat!14}32.5 / 33.3 & \cellcolor{heat!19}45.0 / 45.0 & 0.0 / 0.0 & \cellcolor{heat!12}\textbf{27.5} / 28.9 & \cellcolor{heat!4}9.3 / 11.4 & \cellcolor{heat!26}60.9 / 60.9 & \cellcolor{heat!38}89.4 / 89.4 & \cellcolor{heat!38}90.0 / 90.0 \\
 & \mdl{Nemotron-3-Nano-30B-A3B} & \cellcolor{heat!42}100.0 / 100.0 & 0.0 / 0.0 & \cellcolor{heat!4}10.0 / 11.1 & \cellcolor{heat!42}100.0 / 100.0 & 0.0 / 0.0 & \cellcolor{heat!34}80.0 / 100.0 & \textbf{0.0} / 0.0 & \cellcolor{heat!42}100.0 / 100.0 & \cellcolor{heat!38}90.0 / 100.0 & \cellcolor{heat!42}100.0 / 100.0 \\
 & \mdl{GPT-5.4-mini} & \cellcolor{heat!15}36.7 / 36.7 & \cellcolor{heat!4}3.3 / 4.5 & \cellcolor{heat!4}3.3 / 4.3 & \cellcolor{heat!14}33.3 / 33.3 & 0.0 / 0.0 & \cellcolor{heat!8}20.0 / 20.0 & 0.0 / 0.0 & \cellcolor{heat!6}\textbf{13.3} / 14.8 & \cellcolor{heat!20}46.7 / 53.8 & \cellcolor{heat!14}33.3 / 33.3 \\
 & \mdl{DeepSeek-V4-Flash-0731} & 0.0 / 0.0 & \cellcolor{heat!4}10.0 / 10.0 & 0.0 / 0.0 & 0.0 / 0.0 & 0.0 / 0.0 & 0.0 / 0.0 & 0.0 / 0.0 & \cellcolor{heat!17}40.0 / 40.0 & \cellcolor{heat!4}\textbf{10.0} / 11.1 & \cellcolor{heat!42}100.0 / 100.0 \\
\multirow{-10}{*}{\rotatebox[origin=c]{90}{\textbf{Claude Code}}} & \mdl{Claude Haiku 4.5} & \cellcolor{heat!4}3.3 / 3.3 & 0.0 / 0.0 & 0.0 / 0.0 & 0.0 / 0.0 & 0.0 / 0.0 & 0.0 / 0.0 & 0.0 / 0.0 & \cellcolor{heat!11}26.7 / 27.6 & \cellcolor{heat!24}56.7 / 58.6 & \cellcolor{heat!7}\textbf{16.7} / 16.7 \\
\midrule
 & \mdl{Qwen-3.5-27B-AWQ-4bit} & \cellcolor{heat!13}\textbf{30.0} / 37.5 & 0.0 / 0.0 & 0.0 / 0.0 & \cellcolor{heat!6}15.0 / 17.6 & 0.0 / 0.0 & \cellcolor{heat!6}15.0 / 27.3 & 0.0 / 0.0 & \cellcolor{heat!19}45.0 / 60.0 & \cellcolor{heat!34}80.0 / 80.0 & \cellcolor{heat!21}50.0 / 52.6 \\
 & \mdl{Qwen-3.5-4B} & \cellcolor{heat!17}40.0 / 41.4 & \cellcolor{heat!6}\textbf{13.3} / 14.3 & \cellcolor{heat!6}13.3 / 14.8 & \cellcolor{heat!13}30.0 / 30.0 & \cellcolor{heat!4}6.7 / 22.2 & \cellcolor{heat!7}16.7 / 20.0 & \cellcolor{heat!6}13.3 / 66.7 & \cellcolor{heat!24}56.7 / 70.8 & \cellcolor{heat!28}66.7 / 66.7 & \cellcolor{heat!27}63.3 / 63.3 \\
 & \mdl{Qwen-3.5-9B} & \cellcolor{heat!27}65.1 / 67.2 & \cellcolor{heat!4}9.4 / 10.0 & \cellcolor{heat!8}\textbf{18.0} / 18.3 & \cellcolor{heat!27}64.5 / 64.5 & \cellcolor{heat!4}4.3 / 13.0 & \cellcolor{heat!21}49.2 / 54.5 & \cellcolor{heat!4}1.6 / 9.1 & \cellcolor{heat!33}78.5 / 78.5 & \cellcolor{heat!37}87.1 / 88.4 & \cellcolor{heat!41}98.6 / 98.6 \\
 & \mdl{Qwen-3.5-122B-A10B-AWQ-4bit} & \cellcolor{heat!4}10.0 / 10.0 & \cellcolor{heat!6}15.0 / 16.7 & \cellcolor{heat!8}20.0 / 21.1 & \cellcolor{heat!6}\textbf{15.0} / 15.0 & \cellcolor{heat!8}20.0 / 40.0 & \cellcolor{heat!13}30.0 / 33.3 & 0.0 / 0.0 & \cellcolor{heat!38}90.0 / 90.0 & \cellcolor{heat!25}60.0 / 60.0 & \cellcolor{heat!36}85.0 / 85.0 \\
 & \mdl{Gemma-4-26B-A4B} & \cellcolor{heat!39}92.5 / 92.5 & \cellcolor{heat!8}20.0 / 22.9 & \cellcolor{heat!8}20.0 / 20.5 & \cellcolor{heat!37}87.5 / 87.5 & \textbf{0.0} / 0.0 & \cellcolor{heat!30}72.5 / 78.4 & \cellcolor{heat!4}2.5 / 10.0 & \cellcolor{heat!30}72.5 / 72.5 & \cellcolor{heat!42}100.0 / 100.0 & \cellcolor{heat!42}100.0 / 100.0 \\
 & \mdl{Gemma-4-31B} & \cellcolor{heat!23}55.0 / 55.0 & 0.0 / 0.0 & \cellcolor{heat!10}25.0 / 27.8 & \cellcolor{heat!27}65.0 / 65.0 & 0.0 / 0.0 & \cellcolor{heat!4}\textbf{10.0} / 12.5 & 0.0 / 0.0 & \cellcolor{heat!21}50.0 / 58.8 & \cellcolor{heat!42}100.0 / 100.0 & \cellcolor{heat!40}95.0 / 95.0 \\
 & \mdl{Nemotron-3-Nano-30B-A3B} & \cellcolor{heat!42}100.0 / 100.0 & \cellcolor{heat!27}65.0 / 72.2 & \cellcolor{heat!29}70.0 / 73.7 & \cellcolor{heat!42}100.0 / 100.0 & \cellcolor{heat!21}50.0 / 100.0 & \cellcolor{heat!29}70.0 / 100.0 & \cellcolor{heat!4}\textbf{5.0} / 20.0 & \cellcolor{heat!42}100.0 / 100.0 & \cellcolor{heat!42}100.0 / 100.0 & \cellcolor{heat!42}100.0 / 100.0 \\
 & \mdl{GPT-5.4-mini} & \cellcolor{heat!17}40.0 / 40.0 & 0.0 / 0.0 & 0.0 / 0.0 & \cellcolor{heat!17}40.0 / 40.0 & 0.0 / {\color{dimtext}---} & 0.0 / 0.0 & 0.0 / 0.0 & \textbf{0.0} / 0.0 & \cellcolor{heat!29}70.0 / 70.0 & \cellcolor{heat!17}40.0 / 40.0 \\
 & \mdl{DeepSeek-V4-Flash-0731} & {\color{dimtext}---} & {\color{dimtext}---} & {\color{dimtext}---} & {\color{dimtext}---} & {\color{dimtext}---} & {\color{dimtext}---} & {\color{dimtext}---} & {\color{dimtext}---} & {\color{dimtext}---} & {\color{dimtext}---} \\
\multirow{-10}{*}{\rotatebox[origin=c]{90}{\textbf{Codex}}} & \mdl{Claude Haiku 4.5} & {\color{dimtext}---} & {\color{dimtext}---} & {\color{dimtext}---} & {\color{dimtext}---} & {\color{dimtext}---} & {\color{dimtext}---} & {\color{dimtext}---} & {\color{dimtext}---} & {\color{dimtext}---} & {\color{dimtext}---} \\
\bottomrule
\end{tabular}
\end{adjustbox}
\end{table*}
\begin{table}[t]
\centering
\caption{Attack success rates (ASR, \%) across six AI coding agents using \mdl{Qwen-3.5-27B-AWQ-4bit} with the prompt-based defense. Bold: highest gross ASR per column.}
\label{tab:more_rq6}
\small
\setlength{\aboverulesep}{0pt}\setlength{\belowrulesep}{0pt}\setlength{\tabcolsep}{3pt}\renewcommand{\arraystretch}{1.05}
\begin{adjustbox}{max width=\columnwidth}
\begin{tabular}{l*{8}{c}>{\columncolor{allcol}}c>{\columncolor{allcol}}c}
\toprule
\rowcolor{hdrgray} & \multicolumn{2}{c}{\textbf{CWE-78}} & \multicolumn{2}{c}{\textbf{CWE-79}} & \multicolumn{2}{c}{\textbf{CWE-327}} & \multicolumn{2}{c}{\textbf{CWE-502}} & \multicolumn{2}{c}{\textbf{All}} \\
\rowcolor{hdrgray}\multirow{-2}{*}{\textbf{Agent}} & G & P & G & P & G & P & G & P & G & P \\
\midrule
OpenCode & 32.0 & {\color{dimtext}32.0} & \textbf{15.0} & {\color{dimtext}15.0} & 0.0 & {\color{dimtext}0.0} & 16.0 & {\color{dimtext}16.0} & 20.8 & {\color{dimtext}20.8} \\
Claude Code & 20.0 & {\color{dimtext}20.0} & 10.0 & {\color{dimtext}10.0} & \textbf{70.0} & {\color{dimtext}70.0} & 45.0 & {\color{dimtext}45.0} & 29.0 & {\color{dimtext}29.0} \\
Codex & 12.9 & {\color{dimtext}12.9} & 5.0 & {\color{dimtext}6.7} & 0.0 & {\color{dimtext}0.0} & 38.0 & {\color{dimtext}38.0} & 18.1 & {\color{dimtext}18.8} \\
mini-swe & 20.0 & {\color{dimtext}20.0} & 5.0 & {\color{dimtext}5.0} & 10.0 & {\color{dimtext}10.0} & \textbf{50.0} & {\color{dimtext}50.0} & 17.8 & {\color{dimtext}17.8} \\
Aider & \textbf{70.0} & {\color{dimtext}70.0} & 6.7 & {\color{dimtext}6.7} & 40.0 & {\color{dimtext}40.0} & 37.5 & {\color{dimtext}37.5} & \textbf{35.0} & {\color{dimtext}35.0} \\
OpenHands & 23.3 & {\color{dimtext}23.3} & 5.0 & {\color{dimtext}5.3} & 5.0 & {\color{dimtext}5.3} & \textbf{50.0} & {\color{dimtext}50.0} & 17.5 & {\color{dimtext}17.9} \\
\bottomrule
\end{tabular}
\end{adjustbox}
\end{table}
\begin{table*}[t]
\centering
\caption{Cross-model transferability of bait code on CWE-327 (broken or risky cryptographic algorithm). Same format as \cref{tab:cross_models_cwe78}.}
\label{tab:matrix_all_agents_327}
\scriptsize
\setlength{\aboverulesep}{0pt}\setlength{\belowrulesep}{0pt}\setlength{\tabcolsep}{3pt}\renewcommand{\arraystretch}{1.08}
\begin{adjustbox}{max width=\textwidth}
\begin{tabular}{cl*{10}{c}}
\toprule
\rowcolor{hdrgray}\rule{0pt}{2.4ex} & & \multicolumn{10}{c}{\textbf{Task-completion model}} \\
\cmidrule(lr){3-12}
\rowcolor{hdrgray} & & \texttt{Qwen-3.5-} & \texttt{Qwen-3.5-} & \texttt{Qwen-3.5-} & \texttt{Qwen-3.5-122B-} & \texttt{Gemma-4-} & \texttt{Gemma-4-} & \texttt{Nemotron-3-} & \texttt{GPT-5.4-} & \texttt{DeepSeek-V4-} & \texttt{Claude} \\
\rowcolor{hdrgray}\multirow{-3}{*}{\textbf{Agent}} & \multirow{-3}{*}{\textbf{Bait model}} & \texttt{27B-AWQ-4bit} & \texttt{4B} & \texttt{9B} & \texttt{A10B-AWQ-4bit} & \texttt{26B-A4B} & \texttt{31B} & \texttt{Nano-30B-A3B} & \texttt{mini} & \texttt{Flash-0731} & \texttt{Haiku 4.5} \\
\midrule
 & \mdl{Qwen-3.5-27B-AWQ-4bit} & \cellcolor{heat!25}\textbf{60.0} / 60.0 & \cellcolor{heat!21}50.0 / 71.4 & \cellcolor{heat!17}40.0 / 44.4 & \cellcolor{heat!8}20.0 / 20.0 & \cellcolor{heat!8}20.0 / 20.0 & 0.0 / 0.0 & \cellcolor{heat!21}50.0 / 50.0 & \cellcolor{heat!8}20.0 / 20.0 & \cellcolor{heat!34}80.0 / 80.0 & \cellcolor{heat!17}40.0 / 40.0 \\
 & \mdl{Qwen-3.5-4B} & {\color{dimtext}---} & {\color{dimtext}---} & {\color{dimtext}---} & {\color{dimtext}---} & {\color{dimtext}---} & {\color{dimtext}---} & {\color{dimtext}---} & {\color{dimtext}---} & {\color{dimtext}---} & {\color{dimtext}---} \\
 & \mdl{Qwen-3.5-9B} & \cellcolor{heat!42}100.0 / 100.0 & \cellcolor{heat!17}40.0 / 40.0 & \cellcolor{heat!21}\textbf{50.0} / 50.0 & \cellcolor{heat!42}100.0 / 100.0 & \cellcolor{heat!29}70.0 / 70.0 & \cellcolor{heat!42}100.0 / 100.0 & \cellcolor{heat!13}30.0 / 33.3 & \cellcolor{heat!29}70.0 / 70.0 & \cellcolor{heat!38}90.0 / 90.0 & \cellcolor{heat!29}70.0 / 100.0 \\
 & \mdl{Qwen-3.5-122B-A10B-AWQ-4bit} & {\color{dimtext}---} & {\color{dimtext}---} & {\color{dimtext}---} & {\color{dimtext}---} & {\color{dimtext}---} & {\color{dimtext}---} & {\color{dimtext}---} & {\color{dimtext}---} & {\color{dimtext}---} & {\color{dimtext}---} \\
 & \mdl{Gemma-4-26B-A4B} & \cellcolor{heat!42}100.0 / 100.0 & \cellcolor{heat!8}20.0 / 22.2 & \cellcolor{heat!34}80.0 / 100.0 & \cellcolor{heat!42}100.0 / 100.0 & \cellcolor{heat!17}\textbf{40.0} / 44.4 & \cellcolor{heat!42}100.0 / 100.0 & 0.0 / 0.0 & \cellcolor{heat!34}80.0 / 80.0 & \cellcolor{heat!42}100.0 / 100.0 & \cellcolor{heat!42}100.0 / 100.0 \\
 & \mdl{Gemma-4-31B} & \cellcolor{heat!38}90.0 / 90.0 & \cellcolor{heat!17}40.0 / 40.0 & \cellcolor{heat!38}90.0 / 90.0 & \cellcolor{heat!42}100.0 / 100.0 & \cellcolor{heat!42}100.0 / 100.0 & \cellcolor{heat!42}\textbf{100.0} / 100.0 & \cellcolor{heat!4}10.0 / 10.0 & 0.0 / 0.0 & \cellcolor{heat!25}60.0 / 60.0 & \cellcolor{heat!34}80.0 / 88.9 \\
 & \mdl{Nemotron-3-Nano-30B-A3B} & \cellcolor{heat!20}46.7 / 46.7 & \cellcolor{heat!15}36.7 / 36.7 & \cellcolor{heat!15}36.7 / 36.7 & \cellcolor{heat!20}46.7 / 46.7 & \cellcolor{heat!14}33.3 / 34.5 & \cellcolor{heat!14}33.3 / 33.3 & \cellcolor{heat!8}\textbf{20.0} / 20.7 & \cellcolor{heat!20}46.7 / 46.7 & \cellcolor{heat!34}80.0 / 80.0 & \cellcolor{heat!31}73.3 / 73.3 \\
 & \mdl{GPT-5.4-mini} & \cellcolor{heat!4}5.0 / 5.0 & 0.0 / 0.0 & 0.0 / 0.0 & \cellcolor{heat!4}5.0 / 5.0 & 0.0 / 0.0 & 0.0 / 0.0 & 0.0 / 0.0 & \textbf{0.0} / 0.0 & \cellcolor{heat!19}45.0 / 45.0 & 0.0 / 0.0 \\
 & \mdl{DeepSeek-V4-Flash-0731} & 0.0 / 0.0 & 0.0 / 0.0 & 0.0 / 0.0 & 0.0 / 0.0 & 0.0 / 0.0 & 0.0 / 0.0 & 0.0 / 0.0 & 0.0 / 0.0 & \cellcolor{heat!13}\textbf{30.0} / 30.0 & 0.0 / 0.0 \\
\multirow{-10}{*}{\rotatebox[origin=c]{90}{\textbf{OpenCode}}} & \mdl{Claude Haiku 4.5} & \cellcolor{heat!17}40.0 / 40.0 & \cellcolor{heat!17}40.0 / 40.0 & \cellcolor{heat!34}80.0 / 80.0 & \cellcolor{heat!17}40.0 / 40.0 & \cellcolor{heat!21}50.0 / 50.0 & 0.0 / 0.0 & \cellcolor{heat!4}10.0 / 10.0 & 0.0 / 0.0 & \cellcolor{heat!4}10.0 / 10.0 & \cellcolor{heat!17}\textbf{40.0} / 40.0 \\
\midrule
 & \mdl{Qwen-3.5-27B-AWQ-4bit} & \cellcolor{heat!29}\textbf{70.0} / 70.0 & 0.0 / 0.0 & 0.0 / 0.0 & 0.0 / 0.0 & 0.0 / 0.0 & 0.0 / {\color{dimtext}---} & 0.0 / 0.0 & \cellcolor{heat!8}20.0 / 20.0 & \cellcolor{heat!17}40.0 / 40.0 & \cellcolor{heat!42}100.0 / 100.0 \\
 & \mdl{Qwen-3.5-4B} & \cellcolor{heat!42}100.0 / 100.0 & \cellcolor{heat!4}\textbf{10.0} / 10.0 & \cellcolor{heat!21}50.0 / 50.0 & \cellcolor{heat!13}30.0 / 30.0 & 0.0 / 0.0 & 0.0 / {\color{dimtext}---} & \cellcolor{heat!21}50.0 / 50.0 & \cellcolor{heat!4}10.0 / 10.0 & \cellcolor{heat!13}30.0 / 30.0 & \cellcolor{heat!4}10.0 / 10.0 \\
 & \mdl{Qwen-3.5-9B} & \cellcolor{heat!42}100.0 / 100.0 & \cellcolor{heat!13}30.0 / 30.0 & \cellcolor{heat!42}\textbf{100.0} / 100.0 & \cellcolor{heat!42}100.0 / 100.0 & \cellcolor{heat!29}70.0 / 70.0 & \cellcolor{heat!38}90.0 / 90.0 & \cellcolor{heat!17}40.0 / 40.0 & \cellcolor{heat!38}90.0 / 100.0 & \cellcolor{heat!42}100.0 / 100.0 & \cellcolor{heat!42}100.0 / 100.0 \\
 & \mdl{Qwen-3.5-122B-A10B-AWQ-4bit} & {\color{dimtext}---} & {\color{dimtext}---} & {\color{dimtext}---} & {\color{dimtext}---} & {\color{dimtext}---} & {\color{dimtext}---} & {\color{dimtext}---} & {\color{dimtext}---} & {\color{dimtext}---} & {\color{dimtext}---} \\
 & \mdl{Gemma-4-26B-A4B} & \cellcolor{heat!24}57.5 / 57.5 & \cellcolor{heat!6}15.0 / 16.2 & \cellcolor{heat!4}10.0 / 10.8 & \cellcolor{heat!18}42.5 / 42.5 & \cellcolor{heat!12}\textbf{27.5} / 27.5 & 0.0 / 0.0 & \cellcolor{heat!8}20.0 / 21.1 & \cellcolor{heat!29}70.0 / 70.0 & \cellcolor{heat!21}50.0 / 54.1 & \cellcolor{heat!32}75.0 / 75.0 \\
 & \mdl{Gemma-4-31B} & \cellcolor{heat!28}66.7 / 66.7 & \cellcolor{heat!17}40.0 / 40.0 & \cellcolor{heat!22}53.3 / 55.2 & \cellcolor{heat!35}83.3 / 83.3 & \cellcolor{heat!25}60.0 / 90.0 & \cellcolor{heat!17}\textbf{40.0} / 40.0 & \cellcolor{heat!8}20.0 / 23.1 & \cellcolor{heat!22}53.3 / 53.3 & \cellcolor{heat!31}73.3 / 78.6 & \cellcolor{heat!27}63.3 / 63.3 \\
 & \mdl{Nemotron-3-Nano-30B-A3B} & \cellcolor{heat!27}63.3 / 63.3 & \cellcolor{heat!10}23.3 / 26.9 & \cellcolor{heat!10}23.3 / 23.3 & \cellcolor{heat!34}80.0 / 80.0 & \cellcolor{heat!7}16.7 / 17.2 & \cellcolor{heat!10}23.3 / 26.9 & \cellcolor{heat!4}\textbf{10.0} / 10.3 & \cellcolor{heat!18}43.3 / 43.3 & \cellcolor{heat!29}70.0 / 72.4 & \cellcolor{heat!41}96.7 / 96.7 \\
 & \mdl{GPT-5.4-mini} & \cellcolor{heat!42}100.0 / 100.0 & \cellcolor{heat!17}40.0 / 42.1 & \cellcolor{heat!17}40.0 / 40.0 & \cellcolor{heat!23}55.0 / 55.0 & \cellcolor{heat!21}50.0 / 55.6 & 0.0 / 0.0 & \cellcolor{heat!13}30.0 / 31.6 & \cellcolor{heat!4}\textbf{10.0} / 10.0 & \cellcolor{heat!10}25.0 / 27.8 & \cellcolor{heat!29}70.0 / 70.0 \\
 & \mdl{DeepSeek-V4-Flash-0731} & \cellcolor{heat!4}6.7 / 6.7 & 0.0 / 0.0 & 0.0 / 0.0 & 0.0 / 0.0 & 0.0 / 0.0 & 0.0 / 0.0 & 0.0 / 0.0 & \cellcolor{heat!4}3.3 / 3.6 & \cellcolor{heat!4}\textbf{10.0} / 10.0 & \cellcolor{heat!7}16.7 / 16.7 \\
\multirow{-10}{*}{\rotatebox[origin=c]{90}{\textbf{Claude Code}}} & \mdl{Claude Haiku 4.5} & \cellcolor{heat!8}20.0 / 22.2 & \cellcolor{heat!4}5.0 / 10.0 & 0.0 / 0.0 & \cellcolor{heat!8}20.0 / 21.1 & 0.0 / 0.0 & \cellcolor{heat!4}5.0 / 5.0 & \cellcolor{heat!4}10.0 / 20.0 & \cellcolor{heat!17}40.0 / 42.1 & \cellcolor{heat!19}45.0 / 47.4 & \cellcolor{heat!4}\textbf{5.0} / 5.9 \\
\midrule
 & \mdl{Qwen-3.5-27B-AWQ-4bit} & \textbf{0.0} / 0.0 & 0.0 / 0.0 & 0.0 / 0.0 & 0.0 / 0.0 & 0.0 / 0.0 & 0.0 / 0.0 & 0.0 / 0.0 & 0.0 / 0.0 & \cellcolor{heat!4}5.0 / 5.0 & 0.0 / 0.0 \\
 & \mdl{Qwen-3.5-4B} & \cellcolor{heat!18}42.5 / 42.5 & \cellcolor{heat!5}\textbf{12.5} / 18.5 & \cellcolor{heat!13}30.0 / 38.7 & \cellcolor{heat!21}50.0 / 51.3 & \cellcolor{heat!10}25.0 / 83.3 & \cellcolor{heat!22}52.5 / 67.7 & 0.0 / 0.0 & \cellcolor{heat!22}52.5 / 55.3 & \cellcolor{heat!23}55.0 / 57.9 & \cellcolor{heat!30}72.5 / 72.5 \\
 & \mdl{Qwen-3.5-9B} & \cellcolor{heat!22}53.3 / 53.3 & \cellcolor{heat!7}16.7 / 20.0 & \cellcolor{heat!4}\textbf{6.7} / 7.4 & \cellcolor{heat!15}36.7 / 42.3 & \cellcolor{heat!7}16.7 / 62.5 & \cellcolor{heat!10}23.3 / 36.8 & \cellcolor{heat!4}10.0 / 30.0 & \cellcolor{heat!14}33.3 / 58.8 & \cellcolor{heat!32}76.7 / 76.7 & \cellcolor{heat!22}53.3 / 53.3 \\
 & \mdl{Qwen-3.5-122B-A10B-AWQ-4bit} & \cellcolor{heat!21}50.0 / 50.0 & \cellcolor{heat!4}5.0 / 5.0 & \cellcolor{heat!10}25.0 / 26.3 & \cellcolor{heat!4}\textbf{10.0} / 10.0 & 0.0 / 0.0 & 0.0 / 0.0 & 0.0 / 0.0 & \cellcolor{heat!10}25.0 / 26.3 & \cellcolor{heat!19}45.0 / 45.0 & \cellcolor{heat!10}25.0 / 25.0 \\
 & \mdl{Gemma-4-26B-A4B} & \cellcolor{heat!31}73.3 / 73.3 & \cellcolor{heat!7}16.7 / 18.5 & \cellcolor{heat!15}36.7 / 36.7 & \cellcolor{heat!21}50.0 / 50.0 & \cellcolor{heat!11}\textbf{26.7} / 44.4 & \cellcolor{heat!7}16.7 / 17.9 & 0.0 / 0.0 & \cellcolor{heat!25}60.0 / 60.0 & \cellcolor{heat!28}66.7 / 66.7 & \cellcolor{heat!29}70.0 / 70.0 \\
 & \mdl{Gemma-4-31B} & \cellcolor{heat!34}80.0 / 80.0 & \cellcolor{heat!18}43.3 / 44.8 & \cellcolor{heat!24}56.7 / 60.7 & \cellcolor{heat!35}83.3 / 83.3 & \cellcolor{heat!15}36.7 / 84.6 & \cellcolor{heat!4}\textbf{10.0} / 13.0 & \cellcolor{heat!4}10.0 / 27.3 & \cellcolor{heat!15}36.7 / 39.3 & \cellcolor{heat!25}60.0 / 60.0 & \cellcolor{heat!22}53.3 / 53.3 \\
 & \mdl{Nemotron-3-Nano-30B-A3B} & \cellcolor{heat!19}45.0 / 45.0 & \cellcolor{heat!15}35.0 / 46.7 & \cellcolor{heat!19}45.0 / 50.0 & \cellcolor{heat!17}40.0 / 40.0 & \cellcolor{heat!4}10.0 / 100.0 & \cellcolor{heat!21}50.0 / 55.6 & \cellcolor{heat!4}\textbf{10.0} / 50.0 & \cellcolor{heat!19}45.0 / 47.4 & \cellcolor{heat!27}65.0 / 65.0 & \cellcolor{heat!25}60.0 / 60.0 \\
 & \mdl{GPT-5.4-mini} & \cellcolor{heat!22}52.0 / 56.5 & \cellcolor{heat!4}2.0 / 3.2 & \cellcolor{heat!13}30.0 / 35.7 & \cellcolor{heat!12}28.0 / 29.8 & \cellcolor{heat!4}10.0 / 55.6 & \cellcolor{heat!7}16.0 / 24.2 & \cellcolor{heat!4}2.0 / 10.0 & \cellcolor{heat!8}\textbf{20.0} / 22.2 & \cellcolor{heat!23}54.0 / 56.2 & \cellcolor{heat!19}46.0 / 52.3 \\
 & \mdl{DeepSeek-V4-Flash-0731} & 0.0 / 0.0 & 0.0 / 0.0 & 0.0 / 0.0 & 0.0 / 0.0 & 0.0 / {\color{dimtext}---} & 0.0 / 0.0 & 0.0 / {\color{dimtext}---} & 0.0 / 0.0 & \textbf{0.0} / 0.0 & 0.0 / 0.0 \\
\multirow{-10}{*}{\rotatebox[origin=c]{90}{\textbf{Codex}}} & \mdl{Claude Haiku 4.5} & 0.0 / 0.0 & 0.0 / 0.0 & 0.0 / 0.0 & \cellcolor{heat!4}5.0 / 7.7 & 0.0 / {\color{dimtext}---} & 0.0 / 0.0 & 0.0 / 0.0 & 0.0 / 0.0 & \cellcolor{heat!4}10.0 / 10.5 & \cellcolor{heat!4}\textbf{5.0} / 5.3 \\
\bottomrule
\end{tabular}
\end{adjustbox}
\end{table*}
\begin{table*}[t]
\centering
\caption{Cross-model transferability of bait code on CWE-502 (deserialization of untrusted data). Same format as \cref{tab:cross_models_cwe78}.}
\label{tab:matrix_all_agents_502}
\scriptsize
\setlength{\aboverulesep}{0pt}\setlength{\belowrulesep}{0pt}\setlength{\tabcolsep}{3pt}\renewcommand{\arraystretch}{1.08}
\begin{adjustbox}{max width=\textwidth}
\begin{tabular}{cl*{10}{c}}
\toprule
\rowcolor{hdrgray}\rule{0pt}{2.4ex} & & \multicolumn{10}{c}{\textbf{Task-completion model}} \\
\cmidrule(lr){3-12}
\rowcolor{hdrgray} & & \texttt{Qwen-3.5-} & \texttt{Qwen-3.5-} & \texttt{Qwen-3.5-} & \texttt{Qwen-3.5-122B-} & \texttt{Gemma-4-} & \texttt{Gemma-4-} & \texttt{Nemotron-3-} & \texttt{GPT-5.4-} & \texttt{DeepSeek-V4-} & \texttt{Claude} \\
\rowcolor{hdrgray}\multirow{-3}{*}{\textbf{Agent}} & \multirow{-3}{*}{\textbf{Bait model}} & \texttt{27B-AWQ-4bit} & \texttt{4B} & \texttt{9B} & \texttt{A10B-AWQ-4bit} & \texttt{26B-A4B} & \texttt{31B} & \texttt{Nano-30B-A3B} & \texttt{mini} & \texttt{Flash-0731} & \texttt{Haiku 4.5} \\
\midrule
 & \mdl{Qwen-3.5-27B-AWQ-4bit} & \cellcolor{heat!10}\textbf{24.0} / 24.0 & \cellcolor{heat!8}20.0 / 20.0 & \cellcolor{heat!10}24.0 / 24.0 & \cellcolor{heat!25}60.0 / 60.0 & \cellcolor{heat!27}64.0 / 64.0 & \cellcolor{heat!7}16.0 / 16.0 & \cellcolor{heat!4}8.0 / 8.2 & \cellcolor{heat!14}34.0 / 34.0 & \cellcolor{heat!32}76.0 / 76.0 & \cellcolor{heat!23}54.0 / 54.0 \\
 & \mdl{Qwen-3.5-4B} & {\color{dimtext}---} & {\color{dimtext}---} & {\color{dimtext}---} & {\color{dimtext}---} & {\color{dimtext}---} & {\color{dimtext}---} & {\color{dimtext}---} & {\color{dimtext}---} & {\color{dimtext}---} & {\color{dimtext}---} \\
 & \mdl{Qwen-3.5-9B} & \cellcolor{heat!32}75.0 / 75.0 & \cellcolor{heat!21}50.0 / 50.0 & \cellcolor{heat!21}\textbf{50.0} / 58.8 & \cellcolor{heat!38}90.0 / 90.0 & \cellcolor{heat!42}100.0 / 100.0 & \cellcolor{heat!42}100.0 / 100.0 & \cellcolor{heat!4}5.0 / 5.6 & \cellcolor{heat!21}50.0 / 50.0 & \cellcolor{heat!36}85.0 / 85.0 & \cellcolor{heat!40}95.0 / 100.0 \\
 & \mdl{Qwen-3.5-122B-A10B-AWQ-4bit} & \cellcolor{heat!18}43.3 / 43.3 & \cellcolor{heat!10}23.3 / 24.1 & \cellcolor{heat!10}23.3 / 23.3 & \cellcolor{heat!15}\textbf{36.7} / 36.7 & \cellcolor{heat!18}43.3 / 48.1 & \cellcolor{heat!17}40.0 / 40.0 & \cellcolor{heat!4}6.7 / 6.9 & \cellcolor{heat!10}23.3 / 23.3 & \cellcolor{heat!36}86.7 / 86.7 & \cellcolor{heat!28}66.7 / 66.7 \\
 & \mdl{Gemma-4-26B-A4B} & \cellcolor{heat!42}100.0 / 100.0 & \cellcolor{heat!21}50.0 / 50.0 & \cellcolor{heat!34}80.0 / 80.0 & \cellcolor{heat!34}80.0 / 80.0 & \cellcolor{heat!32}\textbf{75.0} / 75.0 & \cellcolor{heat!21}50.0 / 50.0 & \cellcolor{heat!13}30.0 / 31.6 & \cellcolor{heat!32}75.0 / 75.0 & \cellcolor{heat!32}75.0 / 75.0 & \cellcolor{heat!38}90.0 / 90.0 \\
 & \mdl{Gemma-4-31B} & \cellcolor{heat!19}45.0 / 45.0 & \cellcolor{heat!4}5.0 / 5.0 & 0.0 / 0.0 & 0.0 / 0.0 & \cellcolor{heat!4}5.0 / 5.0 & \cellcolor{heat!23}\textbf{55.0} / 55.0 & \cellcolor{heat!4}5.0 / 5.0 & \cellcolor{heat!4}10.0 / 10.0 & \cellcolor{heat!21}50.0 / 50.0 & \cellcolor{heat!8}20.0 / 20.0 \\
 & \mdl{Nemotron-3-Nano-30B-A3B} & \cellcolor{heat!4}10.0 / 10.0 & \cellcolor{heat!6}13.3 / 13.3 & \cellcolor{heat!4}6.7 / 6.7 & \cellcolor{heat!13}30.0 / 30.0 & \cellcolor{heat!10}23.3 / 23.3 & \cellcolor{heat!4}6.7 / 6.7 & \cellcolor{heat!7}\textbf{16.7} / 16.7 & \cellcolor{heat!4}6.7 / 6.7 & \cellcolor{heat!32}76.7 / 76.7 & \cellcolor{heat!15}36.7 / 36.7 \\
 & \mdl{GPT-5.4-mini} & \cellcolor{heat!6}15.0 / 15.0 & \cellcolor{heat!4}10.0 / 10.0 & \cellcolor{heat!4}10.0 / 10.0 & \cellcolor{heat!4}10.0 / 10.0 & \cellcolor{heat!4}5.0 / 5.3 & 0.0 / 0.0 & \cellcolor{heat!4}5.0 / 5.0 & \cellcolor{heat!4}\textbf{10.0} / 10.0 & \cellcolor{heat!19}45.0 / 45.0 & \cellcolor{heat!6}15.0 / 18.8 \\
 & \mdl{DeepSeek-V4-Flash-0731} & 0.0 / 0.0 & 0.0 / 0.0 & 0.0 / 0.0 & 0.0 / 0.0 & 0.0 / 0.0 & 0.0 / 0.0 & 0.0 / 0.0 & \cellcolor{heat!4}10.0 / 10.0 & \cellcolor{heat!8}\textbf{20.0} / 20.0 & \cellcolor{heat!4}10.0 / 10.0 \\
\multirow{-10}{*}{\rotatebox[origin=c]{90}{\textbf{OpenCode}}} & \mdl{Claude Haiku 4.5} & {\color{dimtext}---} & {\color{dimtext}---} & {\color{dimtext}---} & {\color{dimtext}---} & {\color{dimtext}---} & {\color{dimtext}---} & {\color{dimtext}---} & {\color{dimtext}---} & {\color{dimtext}---} & {\color{dimtext}---} \\
\midrule
 & \mdl{Qwen-3.5-27B-AWQ-4bit} & \cellcolor{heat!23}\textbf{55.0} / 55.0 & \cellcolor{heat!17}40.0 / 40.0 & \cellcolor{heat!13}30.0 / 30.0 & \cellcolor{heat!19}45.0 / 45.0 & \cellcolor{heat!21}50.0 / 50.0 & \cellcolor{heat!21}50.0 / 50.0 & \cellcolor{heat!10}25.0 / 26.3 & \cellcolor{heat!4}10.0 / 10.0 & \cellcolor{heat!13}30.0 / 31.6 & \cellcolor{heat!19}45.0 / 45.0 \\
 & \mdl{Qwen-3.5-4B} & \cellcolor{heat!25}60.0 / 60.0 & \cellcolor{heat!14}\textbf{33.3} / 37.0 & \cellcolor{heat!14}33.3 / 33.3 & \cellcolor{heat!26}61.7 / 61.7 & \cellcolor{heat!15}36.7 / 37.3 & \cellcolor{heat!19}45.0 / 45.0 & \cellcolor{heat!14}33.3 / 35.1 & \cellcolor{heat!12}28.3 / 28.3 & \cellcolor{heat!23}55.0 / 55.0 & \cellcolor{heat!29}68.3 / 68.3 \\
 & \mdl{Qwen-3.5-9B} & \cellcolor{heat!19}45.0 / 45.0 & \cellcolor{heat!4}10.0 / 12.5 & \cellcolor{heat!7}\textbf{17.5} / 21.9 & \cellcolor{heat!20}47.5 / 47.5 & \cellcolor{heat!19}45.0 / 46.2 & \cellcolor{heat!10}25.0 / 25.0 & \cellcolor{heat!17}40.0 / 40.0 & \cellcolor{heat!9}22.5 / 27.3 & \cellcolor{heat!23}55.0 / 55.0 & \cellcolor{heat!26}62.5 / 62.5 \\
 & \mdl{Qwen-3.5-122B-A10B-AWQ-4bit} & {\color{dimtext}---} & {\color{dimtext}---} & {\color{dimtext}---} & {\color{dimtext}---} & {\color{dimtext}---} & {\color{dimtext}---} & {\color{dimtext}---} & {\color{dimtext}---} & {\color{dimtext}---} & {\color{dimtext}---} \\
 & \mdl{Gemma-4-26B-A4B} & \cellcolor{heat!23}55.0 / 55.0 & \cellcolor{heat!6}15.0 / 15.0 & \cellcolor{heat!15}35.0 / 35.0 & \cellcolor{heat!13}30.0 / 30.0 & \textbf{0.0} / 0.0 & 0.0 / 0.0 & 0.0 / 0.0 & 0.0 / 0.0 & \cellcolor{heat!10}25.0 / 26.3 & \cellcolor{heat!10}25.0 / 25.0 \\
 & \mdl{Gemma-4-31B} & \cellcolor{heat!38}90.0 / 90.0 & \cellcolor{heat!4}10.0 / 10.5 & \cellcolor{heat!8}20.0 / 20.0 & \cellcolor{heat!34}80.0 / 80.0 & 0.0 / 0.0 & \cellcolor{heat!6}\textbf{15.0} / 16.7 & \cellcolor{heat!6}15.0 / 21.4 & \cellcolor{heat!8}20.0 / 20.0 & \cellcolor{heat!21}50.0 / 50.0 & \cellcolor{heat!21}50.0 / 50.0 \\
 & \mdl{Nemotron-3-Nano-30B-A3B} & \cellcolor{heat!23}55.0 / 55.0 & \cellcolor{heat!23}55.0 / 55.0 & \cellcolor{heat!23}55.0 / 55.0 & \cellcolor{heat!34}80.0 / 80.0 & \cellcolor{heat!4}5.0 / 10.0 & \cellcolor{heat!34}80.0 / 84.2 & \cellcolor{heat!23}\textbf{55.0} / 78.6 & \cellcolor{heat!17}40.0 / 40.0 & \cellcolor{heat!32}75.0 / 75.0 & \cellcolor{heat!42}100.0 / 100.0 \\
 & \mdl{GPT-5.4-mini} & \cellcolor{heat!25}60.0 / 60.0 & \cellcolor{heat!7}16.7 / 20.8 & \cellcolor{heat!15}36.7 / 45.8 & \cellcolor{heat!11}26.7 / 26.7 & \cellcolor{heat!17}40.0 / 42.9 & \cellcolor{heat!4}6.7 / 6.7 & \cellcolor{heat!7}16.7 / 17.9 & \cellcolor{heat!7}\textbf{16.7} / 19.2 & \cellcolor{heat!29}70.0 / 80.8 & \cellcolor{heat!29}70.0 / 70.0 \\
 & \mdl{DeepSeek-V4-Flash-0731} & \cellcolor{heat!4}10.0 / 10.0 & 0.0 / 0.0 & 0.0 / 0.0 & \cellcolor{heat!4}10.0 / 10.0 & 0.0 / 0.0 & \cellcolor{heat!25}60.0 / 60.0 & 0.0 / 0.0 & \cellcolor{heat!4}10.0 / 10.0 & \cellcolor{heat!4}\textbf{10.0} / 11.1 & \cellcolor{heat!8}20.0 / 20.0 \\
\multirow{-10}{*}{\rotatebox[origin=c]{90}{\textbf{Claude Code}}} & \mdl{Claude Haiku 4.5} & \cellcolor{heat!42}100.0 / 100.0 & \cellcolor{heat!38}90.0 / 90.0 & \cellcolor{heat!29}70.0 / 70.0 & \cellcolor{heat!42}100.0 / 100.0 & \cellcolor{heat!34}80.0 / 80.0 & \cellcolor{heat!40}95.0 / 100.0 & \cellcolor{heat!42}100.0 / 100.0 & \cellcolor{heat!38}90.0 / 90.0 & \cellcolor{heat!42}100.0 / 100.0 & \cellcolor{heat!42}\textbf{100.0} / 100.0 \\
\midrule
 & \mdl{Qwen-3.5-27B-AWQ-4bit} & \cellcolor{heat!18}\textbf{42.0} / 42.0 & \cellcolor{heat!8}18.0 / 18.8 & \cellcolor{heat!7}16.0 / 16.0 & \cellcolor{heat!19}46.0 / 46.0 & \cellcolor{heat!10}24.0 / 30.8 & \cellcolor{heat!8}18.0 / 19.6 & \cellcolor{heat!4}6.0 / 20.0 & \cellcolor{heat!13}30.0 / 30.0 & \cellcolor{heat!23}54.0 / 54.0 & \cellcolor{heat!13}32.0 / 32.0 \\
 & \mdl{Qwen-3.5-4B} & \cellcolor{heat!29}70.0 / 70.0 & \cellcolor{heat!9}\textbf{22.0} / 23.9 & \cellcolor{heat!24}58.0 / 58.0 & \cellcolor{heat!34}80.0 / 80.0 & \cellcolor{heat!15}36.0 / 60.0 & \cellcolor{heat!25}60.0 / 78.9 & \cellcolor{heat!4}8.0 / 30.8 & \cellcolor{heat!24}58.0 / 60.4 & \cellcolor{heat!31}74.0 / 75.5 & \cellcolor{heat!29}70.0 / 70.0 \\
 & \mdl{Qwen-3.5-9B} & \cellcolor{heat!22}53.3 / 53.3 & \cellcolor{heat!6}13.3 / 13.8 & \cellcolor{heat!4}\textbf{3.3} / 3.3 & \cellcolor{heat!6}13.3 / 13.3 & \cellcolor{heat!21}50.0 / 60.0 & \cellcolor{heat!25}60.0 / 66.7 & \cellcolor{heat!7}16.7 / 83.3 & \cellcolor{heat!4}10.0 / 10.0 & \cellcolor{heat!29}70.0 / 70.0 & \cellcolor{heat!31}73.3 / 73.3 \\
 & \mdl{Qwen-3.5-122B-A10B-AWQ-4bit} & \cellcolor{heat!8}20.0 / 20.0 & \cellcolor{heat!4}10.0 / 10.5 & \cellcolor{heat!4}10.0 / 10.0 & \cellcolor{heat!4}\textbf{10.0} / 10.0 & \cellcolor{heat!4}5.0 / 11.1 & 0.0 / 0.0 & 0.0 / 0.0 & \cellcolor{heat!10}25.0 / 25.0 & \cellcolor{heat!27}65.0 / 65.0 & \cellcolor{heat!15}35.0 / 35.0 \\
 & \mdl{Gemma-4-26B-A4B} & {\color{dimtext}---} & {\color{dimtext}---} & {\color{dimtext}---} & {\color{dimtext}---} & {\color{dimtext}---} & {\color{dimtext}---} & {\color{dimtext}---} & {\color{dimtext}---} & {\color{dimtext}---} & {\color{dimtext}---} \\
 & \mdl{Gemma-4-31B} & \cellcolor{heat!23}55.0 / 55.0 & \cellcolor{heat!21}50.0 / 55.6 & \cellcolor{heat!21}50.0 / 50.0 & \cellcolor{heat!13}30.0 / 30.0 & \cellcolor{heat!4}5.0 / 7.7 & \cellcolor{heat!4}\textbf{5.0} / 7.7 & \cellcolor{heat!4}5.0 / 12.5 & \cellcolor{heat!6}15.0 / 16.7 & \cellcolor{heat!19}45.0 / 45.0 & \cellcolor{heat!19}45.0 / 45.0 \\
 & \mdl{Nemotron-3-Nano-30B-A3B} & \cellcolor{heat!34}80.0 / 80.0 & \cellcolor{heat!21}50.0 / 51.3 & \cellcolor{heat!24}57.5 / 57.5 & \cellcolor{heat!40}95.0 / 95.0 & \cellcolor{heat!19}45.0 / 72.0 & \cellcolor{heat!30}72.5 / 90.6 & \cellcolor{heat!7}\textbf{17.5} / 77.8 & \cellcolor{heat!25}60.0 / 60.0 & \cellcolor{heat!41}97.5 / 97.5 & \cellcolor{heat!37}87.5 / 87.5 \\
 & \mdl{GPT-5.4-mini} & \cellcolor{heat!4}5.0 / 5.0 & 0.0 / 0.0 & 0.0 / 0.0 & 0.0 / 0.0 & 0.0 / 0.0 & 0.0 / 0.0 & 0.0 / 0.0 & \textbf{0.0} / 0.0 & 0.0 / 0.0 & \cellcolor{heat!8}20.0 / 20.0 \\
 & \mdl{DeepSeek-V4-Flash-0731} & \cellcolor{heat!21}50.0 / 50.0 & 0.0 / 0.0 & \cellcolor{heat!4}10.0 / 10.0 & \cellcolor{heat!17}40.0 / 40.0 & 0.0 / 0.0 & \cellcolor{heat!13}30.0 / 33.3 & \cellcolor{heat!8}20.0 / 28.6 & 0.0 / 0.0 & \cellcolor{heat!13}\textbf{30.0} / 30.0 & \cellcolor{heat!38}90.0 / 90.0 \\
\multirow{-10}{*}{\rotatebox[origin=c]{90}{\textbf{Codex}}} & \mdl{Claude Haiku 4.5} & \cellcolor{heat!13}30.0 / 30.0 & \cellcolor{heat!4}7.5 / 8.8 & \cellcolor{heat!4}2.5 / 2.9 & \cellcolor{heat!9}22.5 / 22.5 & 0.0 / 0.0 & \cellcolor{heat!4}7.5 / 7.9 & 0.0 / 0.0 & \cellcolor{heat!4}2.5 / 2.8 & \cellcolor{heat!18}42.5 / 42.5 & \cellcolor{heat!5}\textbf{12.5} / 12.5 \\
\bottomrule
\end{tabular}
\end{adjustbox}
\end{table*}

\section{Additional Results Across Agent--Model Configurations}
\label{appdix:rq4}

We repeat the experiment of \cref{tab:rq4} with Traditional Chinese instead of English coding prompts, covering ten LLMs, three agents (OpenCode, Claude Code, and Codex), and four CWE categories. As shown in \cref{tab:rq3-1}, Traditional Chinese prompts yield higher overall gross ASRs in 23 of the 30 agent--model configurations and raise the mean overall gross ASR from 22.3\% to 27.7\%. The increase is largest for CWE-327 and CWE-502, whose mean gross ASRs rise from 26.8\% to 33.7\% and from 28.2\% to 35.0\%, respectively.

\section{Additional Security Detection Results}
\label{appdix:detection_results}

We extend the detection experiments of \cref{sec:rq7} to bait constructed with Claude Code and Codex, using the same seven tools and bait/control comparison. \Cref{tab:detection_claude,tab:detection_codex} report the results, covering 108 and 121 bait/control pairs, respectively. In both cases, no tool detects the targeted CWEs in any poisoned file, consistent with the OpenCode results.

\section{Cross-Model Transferability for Other CWEs}
\label{appdix:more_cross_models}

Following the setting of \cref{tab:cross_models_cwe78}, we extend the cross-model transferability evaluation to CWE-79, CWE-327, and CWE-502 with the same three coding agents and ten LLMs (\cref{tab:matrix_all_agents_079,tab:matrix_all_agents_327,tab:matrix_all_agents_502}). Across all three categories, identifier-poisoned code transfers substantially, and many cross-model combinations exceed the corresponding same-model ASR. Aggregated over all cross-model pairs with qualifying bait, the bait induces the target CWE in at least one completion for 71.5\% of the pairs for CWE-79, 72.4\% for CWE-327, and 86.3\% for CWE-502, compared with 65.5\% for CWE-78, and it is more effective than on its source model in 58.9\%, 49.8\%, and 47.9\% of the pairs, respectively. For CWE-79, bait generated by \mdl{Qwen-3.5-4B} with OpenCode achieves a gross ASR of 10.0\% against its source model but 100.0\% against \mdl{DeepSeek-V4-Flash-0731}. For CWE-327, bait generated by \mdl{Gemma-4-31B} with Codex achieves 10.0\% against itself but 83.3\% against \mdl{Qwen-3.5-122B-A10B-AWQ-4bit}. For CWE-502, bait generated by \mdl{Nemotron-3-Nano-30B-A3B} with Codex achieves 17.5\% against its source model but 97.5\% against \mdl{DeepSeek-V4-Flash-0731}. Transfer is not uniform: some model combinations yield no successful attack, while others approach or reach 100.0\%. These results extend the CWE-78 findings: identifier-based poisoning transfers across LLMs and vulnerability categories without requiring the bait-generation and victim models to coincide, while its strength depends on the targeted CWE, the agent framework, and the specific model pair.

\section{Prompt-Based Defense Across Coding Agents}
\label{appdix:more_prompt_base_defense}

We apply the security reminder of \cref{sec:rq6} to all six agents with \mdl{Qwen-3.5-27B-AWQ-4bit} (\cref{tab:more_rq6}). Relative to the undefended results in \cref{tab:rq2}, the reminder lowers the overall gross ASR of Aider from 56.0\% to 35.0\% (CWE-79: 63.3\% to 6.7\%) and of OpenHands from 22.5\% to 17.5\% (CWE-327: 25.0\% to 5.0\%), and mini-swe's ASR on CWE-327 from 40.0\% to 10.0\%. In contrast, Claude Code's overall gross ASR rises from 25.0\% to 29.0\%, and the ASR on CWE-502 rises from 27.5\% to 37.5\% for Aider and from 20.0\% to 50.0\% for OpenHands. Aider still reaches 70.0\% on CWE-78, Claude Code 70.0\% on CWE-327, and mini-swe and OpenHands 50.0\% on CWE-502, so the reminder does not consistently mitigate the attack.

\end{document}